\documentclass[twocolumn,superscriptaddress,floatfix,preprintnumbers]{revtex4}
\usepackage{graphics,amssymb,amsmath,epsfig,color}
\usepackage{graphicx}

\begin{document}

\title{Full-vortex flux qubit for charged particle optics}
\author{Hiroshi Okamoto}
\affiliation{Department of Electronics and Information Systems, Akita Prefectural University, Yurihonjo 015-0055, Japan}
\date{\today}
\begin{abstract}
We introduce a design of a superconducting flux qubit capable of holding
a full magnetic flux quantum $\phi_{0}$, which arguably is an essential
property for applications in charged particle optics. The qubit comprises
a row of $N$ constituent qubits, which hold a fractional magnetic
flux quantum $\phi_{0}/N$. Insights from physics of the transverse-field
Ising chain reveal that properly designed interaction between these
constituent qubits enables their collective behavior while also maintaining
the overall quantumness. 
\end{abstract}
\maketitle

% main text starts here

\section{Introduction}

Charged particle optics is a potentially fruitful, albeit lesser-known,
application area of quantum science and technology. Among various
proposals \cite{O-L-F,putnam,kruit}, ideas of entanglement-enhanced
electron microscopy (EEEM) \cite{eeem,okamoto-nagatani} and molecule-by-molecule
nano-fabrication \cite{q-interface} have been put forward. These
latter schemes employ superconducting qubits \cite{Superconducting qubits},
which generally produce quantum mechanically superposed electromagnetic
potentials around them. Consequently, a charged particle flying nearby
gets entangled with the qubit. For instance, EEEM would utilize such
entanglement to image fragile biological molecules with a much-needed
signal-to-noise ratio beyond the standard quantum limit under the
condition of a limited allowable number of electrons \cite{looping electron scheme}.

Magnetic flux qubits are preferable to charge qubits especially when
medium or high energy charged particles are used. A closed ring of
magnetic flux quantum $\phi_{0}=h/2e$ is particularly useful \cite{okamoto-nagatani,q-interface}
because the full phase shift $\pi$ is induced on the single-charged
matter wave going through the ring via the Aharonov-Bohm (AB) effect
\cite{AB effect,Tonomura AB}, irrespective of the kinetic energy
or the mass of the particle, while applying effectively zero classical
force to it. A device that naturally comes to mind for generating
a superposition of the presence and absence of a magnetic flux is
the rf-SQUID qubit \cite{rf-SQUID qubit exp}, which can be put in
a quantum mechanically superposed state of two opposing shielding
currents with a suitably adjusted external magnetic field. Furthermore,
a way to make the trapped magnetic flux circular has also been put
forward \cite{okamoto-nagatani}. As shown below, however, on close
examination one finds practical difficulties associated with this
simple idea, despite its soundness at the conceptual level.

\section{Difficulties with a single rf-SQUID\label{sec:Difficulties-with-a}}

We begin by listing several definitions. Let the critical current
of the Josephson junction (JJ), which interrupts the loop inductor
$L$ of an rf-SQUID, be $i_{c}$. The \emph{effective} capacitance,
which include the effect of stray capacitance, is denoted by $C$.
Define $E_{J}=i_{c}\phi_{0}/2\pi$, $E_{C}=\left(2e\right)^{2}/2C$
and $\beta=2\pi Li_{c}/\phi_{0}$. Let the magnetic flux threading
$L$ be $\phi+\phi_{0}/2+\phi_{err}$, where the first term is generated
by the current in the inductor $L$, while $\phi_{0}/2+\phi_{err}$
represents the externally applied bias magnetic flux. Although the
bias magnetic flux is ideally $\phi_{0}/2$, an error $\phi_{err}$
is unavoidable. The potential energy $U\left(\phi\right)=\phi^{2}/2L+E_{J}\cos\left[2\pi\left(\phi+\phi_{err}\right)/\phi_{0}\right]$
has two minima with the difference $\Delta\phi$ (see Fig. 1). Henceforth
we assume $\phi_{err}=0$ unless stated otherwise. It is a unique
character of charged particle optics applications that $\Delta\phi$
should be close to $\phi_{0}$. The potential curve $U\left(\phi\right)$
is shown in Fig. 1. It can be shown that $\Delta\phi\cong\phi_{0}/\left(1+\beta^{-1}\right)$
for $\Delta\phi\cong\phi_{0}$ and hence a large $\beta$ is needed
to keep $\Delta\phi$ close to $\phi_{0}$ (See Appendix A).

Consider EEEM for example \cite{eeem,okamoto-nagatani}. In transmission
electron microscopy (TEM), the lateral size of the coherent electron
wavefront is typically of the order of ${l_{C}\cong10\:\mu\mathrm{m}}$
\cite{Tonomura AB}. The beam divergence in TEM vary, but it can be
less than ${\theta_{d}\cong100\:\mu\mathrm{rad}}$ in experimental
configurations such as Lorenz TEM \cite{Lorenz TEM}. This suggests
that the size of the qubit along the optical axis needs to be smaller
than ${l_{C}/\theta_{d}\cong10\:\mathrm{cm}}$ to keep the beam inside
the qubit, and hence the size should be a few $\mathrm{cm}$ at most.
(The size of the qubit along the axis perpendicular to the optical
axis should be ${l_{C}\cong10\:\mu\mathrm{m}}$.) For rough estimation
purposes, we compute the inductance of the rf-SQUID using the formula
$L\cong\mu_{0}l/2\pi$ for the coaxial cable, neglecting the logarithmic
factor. This results in a value of, e.g., ${L=5\:\mathrm{nH}}$ when
${l\cong2.5\:\mathrm{cm}}$. This value will turn out to be too small. 

We show why a single rf-SQUID qubit does not work in practice. Suppose
that we have $\beta=2\pi Li_{c}/\phi_{0}\cong10\gg1$ to make $\Delta\phi\cong\phi_{0}$.
Then, the critical current of the JJ needs to be ${i_{c}=500\:\mathrm{nA}}$,
assuming the above value ${L=5\:\mathrm{nH}}$. This implies a large
tunnel barrier $\cong2E_{J}\propto i_{c}$ (See Appendix A) between
the two fluxoid states. Hence, a question is whether we have a sufficiently
high rate of tunneling between the two fluxoid states. The state-of-the-art
JJ with ${i_{c}=500\:\mathrm{nA}}$ could have a junction capacitance
$C_{J}$ as small as ${1\:\mathrm{fF}}$. If this is the capacitance
governing the system, our numerical calculation (See Appendix A) gives
an energy splitting of the size ${\Delta\cong1\:\mu\mathrm{eV}}$.
However, $\Delta$ is sensitive to stray capacitance: For example,
having ${C=10\:\mathrm{fF}}$ dramatically suppresses the quantum
tunneling effect, resulting in miniscule energy splitting $\Delta$
smaller than ${30\:\mathrm{pV}}$. (Calculations using the WKB approximation
overestimates $\Delta$.) Such a qubit is not operable because then
${h/\Delta>100\:\mu\mathrm{s}}$ corresponds to e.g. the largest known
decoherence time of superconducting qubits. This lack of robustness
is especially problematic in our case of the large-size rf-SQUID.
Analysis suggests that the loop size of a few cm could result in a
stray capacitance as large as several hundred ${\mathrm{fF}}$ (See
Appendix B).

Qubit decoherence is not the only problem with a miniscule $\Delta$.
This demands a precise alignment of the energy level of the two lowest
potential wells by the external bias flux, because a misalignment
larger than $\Delta$ results in localization of the otherwise symmetric
ground state to one potential well, while the anti-symmetric first
excited state is localized to the other. Such localization is detrimental
to charged particle optics applications \cite{okamoto-nagatani,q-interface}.
The energy difference between the two lowest potential wells is $E_{err}\cong\left(\phi_{0}/L\right)\phi_{err}$
if $\beta\gg1$, when the bias magnetic flux $\phi_{0}/2$ has a small
additional error $\phi_{err}$ (See Appendix A). Since the accuracy
of the bias flux must satisfy ${E_{err}<\Delta<30\:\mathrm{peV}}$,
a necessary condition $\phi_{err}<10^{-9}\phi_{0}$ must be met.

The above condition $\phi_{err}<10^{-9}\phi_{0}$ is difficult to
satisfy. First, a recent experimental study in the context of quantum
computing finds flux noise in a ``coupler loop'', with a $1/f^{0.91}\cong1/f$
form of power spectral density, of the order of $S_{n}=\phi_{n}^{2}/f=\left(10^{-5}\phi_{0}\right)^{2}/f$,
where $f$ is the frequency \cite{MIT annealer}. The variance of
the flux noise in a bandwidth between $f_{L}$ and $f_{H}$ is computed
to be, using a well-known relation, $\int_{\omega_{L}}^{\omega_{H}}S_{n}d\omega/\pi=2\int_{f_{L}}^{f_{H}}\left(\phi_{n}^{2}/f\right)df=\left(2\phi_{n}^{2}/\pi\right)\ln\left(f_{H}/f_{L}\right)\cong\phi_{n}^{2}$,
where we neglected the logarithmic factor and the numerical factor
$2/\pi$ in the last step. The flux fluctuation is thus of the order
of $10^{-5}\phi_{0}$ in this context. A similar value was reported
also in another experiment \cite{Harris coupled rf-SQUIDs}. Second,
analysis of macroscopic resonant tunneling (MRT) allows us to estimate
flux noise both at low and high frequencies \cite{MRT noise measurements}.
In particular, the low frequency studies suggest that the noise is
of the order of $\left(10^{-3}\sim10^{-4}\right)\phi_{0}$ \cite{MRT comment}.
Third, the noise in the output of a SQUID magnetometer, which must
be larger than the magnetic noise in the environment, should therefore
give an upper bound of the environmental noise. (The noise, in terms
of magnetic flux \emph{density}, is typically found smaller with a
larger effective area of the magnetometer \cite{SQUID}, suggesting
that the intrinsic noise of the SQUID plays a role.) A recent study
\cite{SQUID} reports flux noise power density of $\sqrt{S_{n}}=\left(0.09\:\mathrm{fT}/\sqrt{\mathrm{Hz}}\right)\left\{ 1+\left(300\:\mathrm{Hz}/f\right)^{0.3}+\left(3\:\mathrm{Hz}/f\right)\right\} $.
Integrating this from ${f_{L}=0.1\:\mathrm{Hz}}$ to ${f_{H}=10\:\mathrm{GHz}}$
for example, we obtain the variance of magnetic flux noise $\int_{\omega_{L}}^{\omega_{H}}S_{n}d\omega/\pi\cong\left(1\:\mathrm{pT}\right)^{2}$.
Multiplying the aforementioned qubit area of the order of ${l_{C}\times1\:\mathrm{cm}}$,
we obtain the amplitude of the magnetic flux noise of the order of
${10^{-19}\:\mathrm{Wb}\cong10^{-4}\phi_{0}}$, most of which comes
from the frequency independent term. Fourth, if we crudely model the
electromagnetic environment (i.e. the metallic container of the qubit
etc.) as a single inductor $L_{EM}$, there should be thermal magnetic
noise $\phi_{n}$ according to the relation $\phi_{n}^{2}/2L_{EM}\cong k_{B}T/2$.
For example, values ${T=100\:\mathrm{mK}}$ and ${L_{EM}=100\:\mathrm{nH}}$
results in $\phi_{n}\cong0.18\phi_{0}$. Hence the magnetic coupling
between the qubit and $L_{EM}$ must be very small. The above four
findings, when taken together, strongly suggest that $\phi_{err}<10^{-9}\phi_{0}$
is quite unattainable in practice.

\begin{figure}
\includegraphics[scale=0.3]{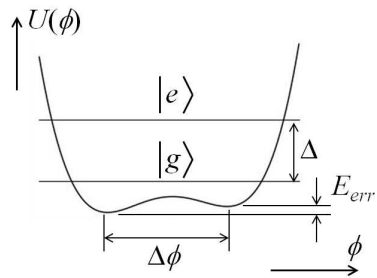}

\caption{Potential curve $U\left(\phi\right)$ of a single rf-SQUID. Energy
levels of the ground state $|g\rangle$ and the excited state $|a\rangle$
are shown for a device with a strong quantum effect, i.e. ${C=10\:\mathrm{fF}}$,
${L=800\:\mathrm{pH}}$ and $\beta=1.11$. The bias energy $E_{err}$
between the wells is exaggerated in the figure. The energy levels
are computed assuming $E_{err}=0$.}

\end{figure}

\section{Proposed solution}

A solution to the above problem is to combine $N>1$ rf-SQUIDs, where
each rf-SQUID is associated with a magnetic flux difference $\Delta\phi=\phi_{0}/N$
between the two fluxoid states. Specifically, we consider $N=4$.
An rf-SQUID with $\beta=\sqrt{2}\pi/4\cong1.11$ has the desired difference
$\Delta\phi=\phi_{0}/4$. (A simple analysis shows that an error of
$1\%$ in $\beta$ corresponds to an error of $4\%$ in $\Delta\phi$.
The expression for $E_{err}$ is modified to be $E_{err}\cong\left(\phi_{0}/4L\right)\phi_{err}$,
when $\Delta\phi=\phi_{0}/4$.) Our strategy of using multiple rf-SQUIDs
may seem simple, but all $N$ SQUIDs should work together, preferably
without using the entire machinery of a universal quantum information
processor. 

For definiteness, let the inductance $L$ of the rf-SQUIDs be ${800\:\mathrm{pH}}$.
This inductance value suggests a size ${l\cong4\:\mathrm{mm}}$ if
we employ the aforementioned formula $L\cong\mu_{0}l/2\pi$. (Further
discussion on estimating $L$ is given at the end of this section.)
Our analysis described in Appendix B suggests that the effective junction
capacitance $C$, mostly coming from stray capacitance, could be as
large as ${60\:\mathrm{fF}}$, but certain measures, such as etching
of silicon inside the inductor loop, $could$ bring this down to about
${3\:\mathrm{fF}}$. Numerical analysis of a single rf-SQUID shows
ample ${\Delta=\left(67\pm3\right)\:\mu\mathrm{eV}}$ if ${C=10\:\mathrm{fF}}$
(see Appendix A, where we assumed $1\%$ uncertainty in the values
of $\alpha$ and $\beta$). Figure 1 shows this case. In the case
of ${C=100\:\mathrm{fF}}$, we obtain ${\Delta=\left(7.9\pm0.5\right)\:\mu\mathrm{eV}}$
(again with $1\%$ uncertainty in $\alpha$ and $\beta$), which may
still be acceptable and shows certain degree of robustness of the
energy splitting $\Delta$. (Possibly, a moderately large $C$ may
even be advantageous because then the flux $\phi$ is well-localized.
See Sec. V along with Appendix E for the effect of quantum mechanically
uncertain $\phi$.) 

In practice, device parameters vary from one JJ to another. To achieve
small spread in device parameters among multiple rf-SQUIDs, the use
of the compound JJ (CJJ), which is effectively a JJ with adjustable
$i_{c}$, will likely be necessary \cite{compound JJ}. To uniformly
modulate the parameters of multiple rf-SQUIDs during operation (see
Sec. VI), the more complex compound-CJJ (CCJJ) may be needed \cite{compound-compound JJ}.
In the rest of this paper, the term ``JJ'' will mean \emph{effective}
JJ that may actually be CJJ or CCJJ. 

We consider a 1-dimensional (1D) chain of $N=4$ rf-SQUIDs, along
which charged particles fly. These rf-SQUIDs work together as a single
qubit, which we will call the \emph{composite qubit}. For the moment,
we regard each rf-SQUID as a spin $1/2$, labeled consecutively as
$k=1,2,\cdots,N$. Let the $k$-th spin's basis states be $\mid\uparrow\rangle_{k}$
and $\mid\downarrow\rangle_{k}$, which correspond to the two fluxoid
states of the $k$-th rf-SQUID. Define symmetric and antisymmetric
states respectively as $|s\rangle_{k}=\left(\mid\uparrow\rangle_{k}+\mid\downarrow\rangle_{k}\right)/\sqrt{2}$
and $|a\rangle_{k}=\left(\mid\uparrow\rangle_{k}-\mid\downarrow\rangle_{k}\right)/\sqrt{2}$.
For charged particle optics applications, the ground state of noninteracting
spins $\otimes_{k=1}^{N}|s\rangle_{k}$ is useless. The basis states
of the composite qubit should instead be the ``ferromagnetic'' $\mid\Uparrow\rangle=\otimes_{k=1}^{N}\mid\uparrow\rangle_{k}$
and $\mid\Downarrow\rangle=\otimes_{k=1}^{N}\mid\downarrow\rangle_{k}$.
We will show that suitable ferromagnetic interaction between the spins
gives what we want. The low-lying energy eigenstates should essentially
be $\left(\mid\Uparrow\rangle\pm\mid\Downarrow\rangle\right)/\sqrt{2}$
because of tunneling between the states $\mid\Uparrow\rangle$ and
$\mid\Downarrow\rangle$, which does occur since $N$ is finite in
our case. Many methods to couple flux qubits have been studied and
demonstrated, including tunable ones \cite{JJ coupling}.

\begin{figure*}
\includegraphics[scale=0.2]{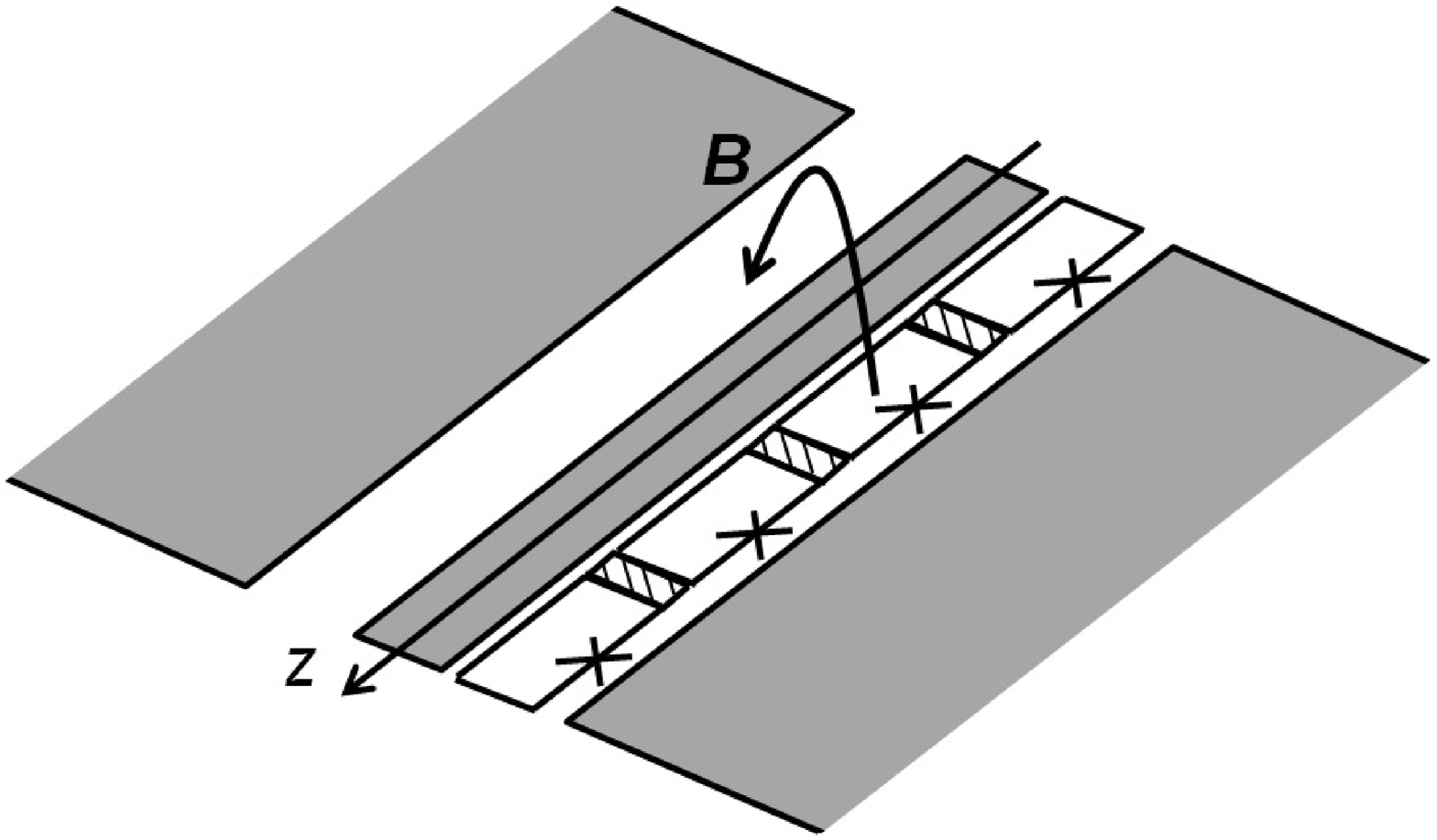}(a)\hspace{5em} \includegraphics[scale=0.2]{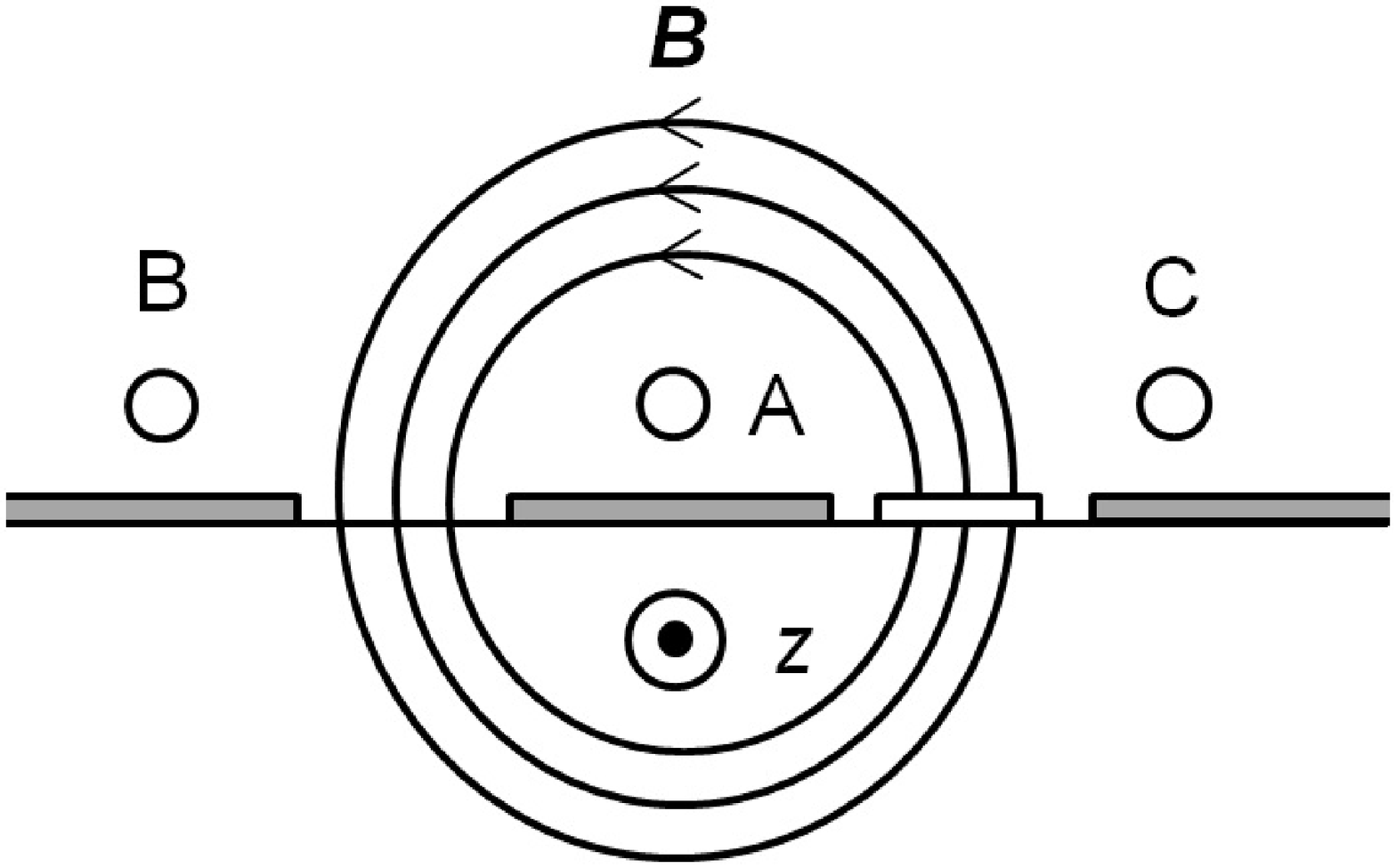}(b)

\caption{(a) Proposed device structure comprising a row of $N=4$ rf-SQUIDs
on a surface of a substrate. Each ``cross'' symbol represents a
\emph{compound} JJ in an abbreviated way. Charged particles fly in
close proximity to, and in parallel with, the surface along the $z$
axis. Superconducting planes (shaded parts) are placed so that the
field lines of magnetic flux density $\mathbf{\boldsymbol{\mathbf{\mathit{B}}}}$
are forced to make a loop. Adjacent rf-SQUIDs interact through couplers,
which are represented by hatched blocks. Flux biasing coils, possibly
using persistent current trapping \cite{persistent current trapping},
are not shown. (b) The cross section of the device, perpendicular
to the optical axis $z$. The charged particle beam A goes through
the magnetic flux ring, whereas the beams B and C do not. The white
box represents an rf-SQUID, while shaded boxes represent superconducting
films. The shape of the field lines is only schematic. See the main
text for further discussions on the shape of the magnetic flux ring.}
\end{figure*}

Figure 2 illustrates a possible implementation of the composite qubit.
A row of $4$ rf-SQUIDs are placed on a substrate as shown in Fig.
2 (a). Charged particles fly along the optical $z-$axis. Each rf-SQUID
is biased with an external flux $\phi_{0}/2$, either by an additional
coil or by persistent-current-trapping \cite{persistent current trapping}.
Either way, care should be taken to avoid known difficulties in biasing
an rf-SQUID qubit \cite{Ralph et al,Flux biasing difficulties} and
known methods should be employed as needed to take care of cross coupling
between bias controls \cite{JJ coupling}. Adjacent rf-SQUIDs interact
ferromagnetically through a coupling circuit. These rf-SQUIDs are
arranged with suitably grounded superconducting strips in such a way
that the magnetic field from each rf-SQUID primarily makes a loop
shown. The geometric design should be such that the field does not
go sideways, i.e. to the adjacent rf-SQUIDs. At the same time, these
superconducting strips, especially the one on the side of JJs, should
be carefully designed so that they do not contribute much stray capacitance.
Figure 2 (b) shows a cross section, perpendicular to the $z-$axis,
of the device. A nominal phase difference $0$ or $\pi$ is produced
between the charged particle wave going through the magnetic flux
ring (A) and the waves passing by the ring (B and C), depending on
the qubit state. These beams A, B, and C may be generated using a
stencil mask in the upstream of the charged particle beam. It should
be easy to envision using techniques in the field of microelectromechanical
systems (MEMS) to, for example, make a groove where the beam A goes,
etc.

Further consideration is warranted on the estimation of the loop inductance
$L$. The magnetic energy stored in an inductor is 
\begin{equation}
\frac{1}{2\mu_{0}}\int\boldsymbol{B}^{2}dV=\frac{1}{2L}\left\{ \int\boldsymbol{B}\cdot d\boldsymbol{S}\right\} ^{2},
\end{equation}
in the case of a single-turn coil without magnetic material. Hence
$L=\mu_{0}\left\{ \int\boldsymbol{B}\cdot d\boldsymbol{S}\right\} ^{2}/\int\boldsymbol{B}^{2}dV$,
in which the magnetic flux density$\boldsymbol{B}$ may be replaced
with any vector field that is proportional to $\boldsymbol{B}$, is
determined entirely by the \emph{shape} of magnetic field lines. For
example, one may use the inductance formula for coaxial cables if
the magnetic flux lines have the tight tube-like shape. Such a shape
can in principle be formed by suitably placing coaxial superconducting
tubes around the rf-SQUID. However, an application at hand may allow
for more extended magnetic field distributions. For the example shown
in Fig. 2 (b), the electron beams are only at positions A, B, and
C and the magnetic field lines do not have to be tightly held together
above the beam positions. In such cases, $L$ is larger for a similar
size of rf-SQUIDs. As an extreme example, numerical analysis using
InductEX software shows that a superconducting rectangular loop, with
the width of ${10\:\mu\mathrm{m}}$ containing magnetic flux, on a
flat substrate without any other superconducting part, has an inductance
per unit length of $\cong{1.2\:\mathrm{nH/mm}}$ \cite{inductEX}.
This value is significantly larger than the case of a coaxial cable,
where $\mu_{0}l/2\pi={0.2\:\mathrm{nH/mm}}$.

\section{The Lagrangian}

We model ferromagnetic interaction between neighboring rf-SQUIDs produced
by the coupler circuits. For simplicity, we ignore boundary effects
at both the ends of the chain. Let $L$ and $M$ respectively be the
self inductance common to all the rf-SQUIDs and the effective mutual
inductance common to all the neighboring pairs of rf-SQUIDs. The magnetic
flux in the $k-$th rf-SQUID is $\phi_{k}+\phi_{0}/2$, where $\phi_{0}/2$
is the bias flux. Under a condition $M/L\ll1$, a straightforward
analysis (See Appendix C) shows that the magnetic energy stored in
the system is $U_{mag}\cong\left[1/2\left(L+2M\right)\right]\sum_{k=1}^{N}\phi_{k}^{2}+\left(M/2L^{2}\right)\sum_{k=1}^{N-1}\left(\phi_{k+1}-\phi_{k}\right)^{2}$,
to the first order in $M/L$. Henceforth we will use the above expression
as if it is exact. To set up the Lagrangian $\hat{L}$ of the system,
we define $\theta_{k}=2\pi\phi_{k}/\phi_{0}$, the gauge-invariant
phase difference across the $k$-th JJ plus $\pi$, which we take
as dynamical variables. The charging energy of the JJ capacitance
gives the kinetic energy because it involves $\dot{\theta}_{k}=d\theta_{k}/dt$.
On the other hand, the potential energy is stored in the inductors
and the JJs. We obtain \begin{widetext} 

\begin{equation}
\hat{L}=\sum_{k=1}^{N}\left\{ \frac{C\phi_{0}^{2}}{2}\left(\frac{\dot{\theta}_{k}}{2\pi}\right)^{2}-\frac{M\phi_{0}^{2}}{2L^{2}}\left(\frac{\theta_{k+1}-\theta_{k}}{2\pi}\right)^{2}-E_{J}\cos\theta_{k}-\frac{\phi_{0}^{2}}{2\left(L+2M\right)}\left(\frac{\theta_{k}}{2\pi}\right)^{2}\right\} .\label{eq:Lagrangian}
\end{equation}
\end{widetext} 

The Lagrangian (\ref{eq:Lagrangian}) equivalently describes coupled
inverted mechanical pendulums with a restoration force (the last term).
We use this mechanical analog to aid our intuition. These pendulums
are independent without interaction ($M/L\rightarrow0$). Discretizing
the quantum state space, we say that if the $k$-th pendulum is in
the stable state $\theta_{k}<0$, then it is in the down state $\mid\downarrow\rangle_{k}$
and likewise the stable $\theta_{k}>0$ corresponds to $\mid\uparrow\rangle_{k}$.
Note that all pendulums would effectively act as a single object and
the ground state would be the desired entangled state $\left(\mid\Uparrow\rangle+\mid\Downarrow\rangle\right)/\sqrt{2}$
if the couplings among the pendulums are sufficiently strong. However,
the mass and the energy barrier height for the group of pendulums
are $N$ times those of the individual pendulum and quantum tunneling
would be strongly suppressed. Consequently, a problem arises as to
whether the decoherence time is longer than the time scale associated
with the energy splitting and if the required accuracy of the bias
magnetic flux is attainable. Our central question is whether there
exists an intermediate coupling strength, where both the entangled
ground state and sufficiently strong quantum fluctuation are realized.

\section{Many-body physics governing the system}

Instead of calculating properties of the set of $4$ rf-SQUIDs by
brute-force, we intend to gain broad physical insights from known
many-body physics. Hence, despite that the number of rf-SQUIDs we
consider is only $N=4$, we approximate it by infinity. We regard
the composite qubit as a set of interacting spins. Our strategy is
to see how the system changes upon renormalization: If the ``quantumness''
is kept upon renormalization, then we have evidence that $N$ rf-SQUIDs
as a whole, or less precisely the ``single renormalized rf-SQUID'',
would keep desired overall quantumness. The ``bare'' parameters
in renormalization theory correspond to the device parameters of the
individual rf-SQUID. 

Our model is described by the Hamiltonian \cite{Harris coupled rf-SQUIDs}
\begin{equation}
H=-J\sum_{i}\sigma_{i}^{z}\sigma_{i+1}^{z}-h\sum_{i}\sigma_{i}^{x}-\varepsilon\sum_{i}\sigma_{i}^{z},\label{eq:extended transverse Ising}
\end{equation}
where $\sigma$s are the Pauli matrices pertaining to the spins discussed
above. This is the 1D transverse-field Ising model if $\varepsilon=0$,
which has been extensively studied as a prototypical model to study
quantum phase transitions \cite{Transverse Ising RG} and has also
attracted much attention recently in the context of quantum annealing
\cite{quantum annealing}. The last term in Eq. (\ref{eq:extended transverse Ising}),
which is assumed to be small, is added to account for non-ideal flux
biasing. Appropriate assignments of the variables are seen to be $J=M\phi_{0}^{2}/2N^{2}L^{2}$,
$h=\Delta/2$ and $\varepsilon=E_{err}/2$ to map to the rf-SQUID
case. The massive sine-Gordon model may seem more accurate, but aside
from solvability the classical soliton size $\cong\sqrt{J/E_{J}}$
turns out to be small in the parameter region of interest, justifying
the use of a spin-based model.

First, we examine the overall behavior. It is known that the ground
state is ferromagnetic or paramagnetic if $R=h/J<1$ or $R>1$ respectively
\cite{Transverse Ising RG}. We note that a similar \emph{superposed}
ferromagnetic state has been observed experimentally in a system comprising
$8$ flux qubits \cite{PRX experiment}. It is also known that $R$
evolves according to a simple rule $R'=R^{2}$ upon block renormalization
of 2 spins \cite{Fernandez-Pacheco}. This rule is robust as long
as $\varepsilon$ is small (See Appendix D). This shows that the overall
system behaves similarly to the constituent spins, or the rf-SQUIDs,
if the system is near the quantum critical point (QCP) $R=1$. We
introduce $\kappa=R-1$ to indicate the distance from the QCP.

The composite qubit must have two basis states by definition. The
basis states $\mid\Uparrow'\rangle$ and $\mid\Downarrow'\rangle$
should respectively be similar to the totally polarized states$\mid\Uparrow\rangle$
and $\mid\Downarrow\rangle$, in the sense that they induce well-defined
phase shifts differing by $\pi$ to the charged particle wave. A natural
condition for any such basis state $|b\rangle$ is that the 2-point
spin correlator $C\left(n\right)=\langle b|\sigma_{i}^{z}\sigma_{i+n}^{z}|b\rangle$
is close to $1$, for all positive $n$ smaller than the size of the
spin chain $N$. Conversely, if a state satisfies this condition,
then the state must be similar to either $\mid\Uparrow\rangle$, $\mid\Downarrow\rangle$
or their superposition. It is known in the case of an infinite chain
at $\varepsilon=0$ that, at zero temperature (i.e. with a ground
state) and also in the limit $n\rightarrow\infty$, $C\left(n\right)=\left(2\pi^{2}n^{2}\kappa\right)^{-1/4}e^{-n\kappa}$
if $\kappa>0$ (the paramagnetic phase) and $C\left(n\right)=\left(2\left|\kappa\right|\right)^{1/4}$
if $\kappa<0$ (the ferromagnetic phase) \cite{Transverse Ising RG}.
Under the condition $\kappa<0$, the polarization of spins over the
\emph{entire} infinite chain is $\langle b|\sigma_{i}^{z}|b\rangle=\sqrt{C\left(n\right)}=\left(2\left|\kappa\right|\right)^{1/8}$.
However, we expect stronger polarization over a finite length, especially
when the length is shorter than the average size of the ``magnetic
domain''. In the paramagnetic region $\kappa>0$, no polarization
over the entire infinite chain is present. However, within a finite
distance we still have polarization, e.g. $C\left(n\right)\varpropto\kappa^{1/4}$
at $n\cong1/\kappa$, and hence our finite system could essentially
be fully polarized for a small enough $\kappa$. Hence, the composite
qubit might work also in the $\kappa>0$ region.

Despite the above remark on stronger polarization over a finite length,
here we proceed conservatively. We require a $\kappa$ value, corresponding
to the polarization $P$ of the \emph{infinite} chain, to be sufficiently
close to the full value $1$. From the perspective of charged particle
optics, the two basis states of the composite qubit should have magnetic
flux difference close to $\phi_{0}$. This does not necessarily mean
that $P$ needs to be close to $1$ because the magnetic flux difference
can also be adjusted by modestly varying $\beta$. (For example, $\beta=2\pi/3\sqrt{3}\cong1.21$
would give the nominal flux difference $4\phi_{0}/3$.) However, a
small value of $P$ generally means larger quantum uncertainty in
the value of magnetic flux, which entails unwanted entanglement with
a charged particle that could lead to excitation of the composite
qubit. Although evaluation of such effects is a complex problem that
is beyond the scope of the present work (See Appendix E for a preliminary
analysis for EEEM), it is reasonable to assume that a value of $P$
close to $1$ should limit the size of aforementioned quantum uncertainty.
Hence, for the sake of rough estimation, we assume $P=\left|\kappa\right|^{1/8}$
so that we obtain the full polarization $P=1$ at $\kappa=-1$, i.e.
$R=0$. This conservative relation underestimates the known polarization
$\langle b|\sigma_{i}^{z}|b\rangle=\left(2\left|\kappa\right|\right)^{1/8}$
near the QCP. For example, values of $P=0.90$ and $0.80$ respectively
correspond to values of $R=0.57$ and $0.83$ because of the relation
$R=1-P^{8}$. 

To be specific, we analyze the case of $R<1$. We slightly extend
the block renormalization scheme \cite{Fernandez-Pacheco} for the
transverse-field Ising model to the case where a weak but non-zero
longitudinal field is present (See Appendix D). Upon replacing each
block of 2 spins with a renormalized spin, renormalized Hamiltonian
parameters are obtained as 
\[
J'=\frac{1}{\sqrt{1+R^{2}}}J,\:h'=\frac{R}{\sqrt{1+R^{2}}}h,
\]
\begin{equation}
\varepsilon'=\left\{ 1+\frac{1+2R^{2}}{\left(1+R^{2}\right)^{3/2}}\right\} \varepsilon.\label{eq:renormalization rule}
\end{equation}
To be on the safe side, we assumed that the flux biasing errors $\varepsilon$
are with the same sign and magnitude for all rf-SQUIDs, although in
reality we expect random biasing errors. To maintain quantumness,
$h$ should not decrease too quickly upon renormalization in the region
$h<J$. Although making $h$ close to $J$ minimizes the rate of decrease,
this entails a smaller polarization $P$ and hence a compromise should
be made. At the same time, $\varepsilon$ should not grow to exceed
$h$ because of the analogous relation $E_{err}<\Delta$ in the case
of the rf-SQUID. Equations (\ref{eq:renormalization rule}) imply
that both $J$ and $h$ grow proportional to $1/\sqrt{N}$ compared
to the initial values, and similarly $\varepsilon\propto N^{\ln\left[\left(3+2\sqrt{2}\right)/2\sqrt{2}\right]/\ln2}\cong N^{1.0}$
at the QCP.

Numerical calculations away from the QCP show the followings (See
Appendix F). After $2$ iterations of block renormalization, implying
that $N=2^{2}=4$ spins are combined to make a renormalized spin,
the parameters $h$ and $\varepsilon$ evolve into renormalized values
$h''$ and $\varepsilon''$ that depend on the initial value of $R$.
For the aforementioned two values $R=0.57$, $0.83$ and the QCP value
$R=1.0$, we respectively obtain $\left(h''/h\right)/\left(\varepsilon''/\varepsilon\right)=0.036$,
$0.083$ and $0.12$. Hence the ratio $h/\varepsilon$ decreases by
$1\sim2$ orders of magnitude upon renormalization, implying that
the original, constituent rf-SQUIDs must satisfy $10^{2\sim3}E_{err}\cong\Delta$
in order to have a margin of an order of magnitude. For the parameters
mentioned before, i.e. ${L=800\:\mathrm{pH}}$, ${C=10\:\mathrm{fF}}$,
and $\beta=1.11$, the bias flux error must satisfy $\phi_{err}<10^{-\left(4\sim5\right)}\phi_{0}$.
The requirement will be an order of magnitude more stringent in the
case of larger stray capacitance ${C=100\:\mathrm{fF}}$. Discussions
in Sec. \ref{sec:Difficulties-with-a} suggests that the requirement
$\phi_{err}<10^{-\left(4\sim5\right)}\phi_{0}$ does not seem to be
out of the realm of feasibility, although we must strive to minimize
stray capacitance. Nonetheless, we have consistently been on the safe
side and hence the biasing precision requirement may well be relaxed.
Also note that the precision requirement is exponentially harder,
and hence virtually impossible to satisfy, if we use a single large
rf-SQUID qubit instead of the proposed composite qubit. Since the
rf-SQUID coil has the inductance ${L=800\:\mathrm{pH}}$ with the
size ${l\cong4\:\mathrm{mm}}$ using the aforementioned formula $L\cong\mu_{0}l/2\pi$,
the entire size of $N=4$ rf-SQUIDs is ${4l\cong16\:\mathrm{mm}}$.
This is well within the required length of a few $\mathrm{cm}$ mentioned
in Sec \ref{sec:Difficulties-with-a}. 

\section{Operation of the composite qubit}

Both in EEEM and non-invasive charge detection, we start with resetting
the qubit in the state $\left(\mid\Uparrow'\rangle+\mid\Downarrow'\rangle\right)/\sqrt{2}$.
This is done by, first setting all the rf-SQUID in the symmetric ground
state when they are uncoupled at a large $R$. Then we go through
the QCP adiabatically and reach the operation point such as $R=0.57$.
In the infinite chain, the excitation gap is known to vanish at the
QCP \cite{Transverse Ising RG} and the ground level is degenerate
when $R<1$, but our system is finite. In order to roughly estimate
the energy difference between the ground state and the excited state
of the qubit, we use the renormalized version of $\Delta=2h$, which
is the $2$-times renormalized parameter $\Delta''=2h''=0.31h=0.15\Delta={1.6\times10^{-24}\:\mathrm{J}}$
(See Table AI of Appendix A), where we used the value $h''/h=0.153$
at $R=0.57$ (See Appendix F). Hence we need to take the time larger
than $h/\Delta''={0.40\:\mathrm{ns}}$ to reach the operation point,
where the qubit interacts with the charged particles. Then we either
wait, so that the qubit state rotates about the axis of the Bloch
sphere connecting the points $\left(\mid\Uparrow'\rangle\pm\mid\Downarrow'\rangle\right)/\sqrt{2}$,
or apply a tiny additional flux to the biasing flux to rotate the
state about the axis connecting $\mid\Uparrow'\rangle$ and $\mid\Downarrow'\rangle$
\cite{non-microwave control of qubits}. Finally, the composite qubit
is measured, for example magnetically, with respect to the basis states
$\mid\Uparrow'\rangle$ and $\mid\Downarrow'\rangle$. The entire
operation should be carried out within the decoherence time of the
composite qubit.

\section*{ACKNOWLEDGMENT}

In one of earlier versions of the manuscript, the author proposed
the use of multi-turn rf-SQUID loops. An anonymous referee pointed
out, providing numerical evidence, that such a scheme would be unnecessary.
This valuable and important observation, together with the problem of stray capacitance,
prompted the author to go back to the initial idea of using rf-SQUIDs
with $\beta\gtrsim1$. This research was supported in part by the
JSPS Kakenhi (grant No. 25390083).

\appendix
\renewcommand\thefigure{\thesection\arabic{figure}}
\renewcommand\thetable{\thesection\Roman{table}}

\section{ANALYSIS OF A SINGLE RF-SQUID}

\setcounter{figure}{0} Differentiating the potential function $U\left(\phi\right)=\phi^{2}/2L+E_{J}\cos\left(2\pi\phi/\phi_{0}\right)$
and equating the result to zero, we obtain 
\begin{equation}
\frac{\pi\Delta\phi}{\phi_{0}}=\beta\sin\left(\frac{\pi\Delta\phi}{\phi_{0}}\right),\label{eq:minima condition}
\end{equation}
because the two potential minima are at $\pm\Delta\phi/2$. Setting
$\Delta\phi=\phi_{0}/4$, which corresponds to the composite qubit
comprising $N=4$ rf-SQUIDs, results in $\beta=\sqrt{2}\pi/4$. 

If we were to use a single rf-SQUID with $\Delta\phi\lesssim\phi_{0}$,
we obtain 
\[
\frac{\pi\Delta\phi}{\phi_{0}}=\beta\sin\left(\frac{\pi\Delta\phi}{\phi_{0}}\right)=\beta\sin\left\{ \frac{\pi\left(\phi_{0}-\Delta\phi\right)}{\phi_{0}}\right\} 
\]
\begin{equation}
\cong\beta\frac{\pi\left(\phi_{0}-\Delta\phi\right)}{\phi_{0}},
\end{equation}
resulting in the expression $\Delta\phi\cong\phi_{0}/\left(1+\beta^{-1}\right)$
mentioned in the main text. This, however, entails a large energy
barrier: 
\begin{equation}
U\left(0\right)-U\left(\frac{\Delta\phi}{2}\right)\cong2E_{J}\left(1-\frac{\pi^{2}}{4\beta}\right)\cong2E_{J}
\end{equation}

The energy splitting $\Delta$ between the ground state $|g\rangle$
and the first excited state $|a\rangle$ is computed by solving the
Schrodinger equation numerically. (It turns out that the semiclassical
WKB method is not applicable.) Let $E_{L}$ be $\left(\phi_{0}/2\pi\right)^{2}/2L$.
We introduce a dimensionless Hamiltonian $\hat{h}=H/E_{C}$, where
$H$ is the Hamiltonian. This can be written as, in the ``position
representation'' 
\begin{equation}
\hat{h}=-\frac{\partial^{2}}{\partial\theta^{2}}+\alpha\theta^{2}+2\alpha\beta\cos\theta,
\end{equation}
where $\theta=2\pi\phi/\phi_{0}$ and $\alpha=E_{L}/E_{C}$. We then
use the standard Numerov method to obtain eigenvalues of $\hat{h}$.

The energy difference $E_{err}$ between the two potential minima
is evaluated using the potential curve $U\left(\phi\right)=\phi^{2}/2L+E_{J}\cos\left[2\pi\left(\phi+\phi_{err}\right)/\phi_{0}\right]$,
where $\phi_{err}\ll\phi_{0}$ is a small error in the bias flux.
Since 
\begin{equation}
U\left(\phi\right)\cong\left[U\left(\phi\right)\right]_{\phi_{err}=0}-i_{c}\phi_{err}\sin\left(\frac{2\pi\phi}{\phi_{0}}\right),
\end{equation}
 quantum mechanically one may treat the second term as perturbation
to obtain 
\begin{equation}
E_{err}\cong i_{c}\phi_{err}\left\{ \langle\uparrow|\sin\left(\frac{2\pi\phi}{\phi_{0}}\right)|\uparrow\rangle-\langle\downarrow|\sin\left(\frac{2\pi\phi}{\phi_{0}}\right)|\downarrow\rangle\right\} ,
\end{equation}
where $|\uparrow\rangle=\left(|g\rangle+|e\rangle\right)/\sqrt{2}$
and $|\downarrow\rangle=\left(|g\rangle-|e\rangle\right)/\sqrt{2}$
are states localized in either of the two potential wells for the
unperturbed system. If classical approximation is valid, then the
wave packet is well-localized at a potential minimum and we obtain
\begin{equation}
E_{err}\cong2i_{c}\sin\left(\frac{\pi\Delta\phi}{\phi_{0}}\right)\cdot\phi_{err}=\frac{\Delta\phi}{L}\phi_{err},
\end{equation}
where we used Eq. (\ref{eq:minima condition}) in the second equality.
(Alternatively, one can compute $E_{err}\cong U\left(\phi_{-}\right)-U\left(\phi_{+}\right)$,
where the two minima $\phi_{\pm}$ are evaluated to the first order
in $\phi_{err}$. This conceptually simpler method gives the same
result.) Hence follow the expressions in the main text, namely $E_{err}\cong\left(\phi_{0}/4L\right)\phi_{err}$
for the composite qubit and $E_{err}\cong\left(\phi_{0}/L\right)\phi_{err}$
for the single rf-SQUID. One should keep in mind that these estimates
use the semi-classical approximation. 

To find the effect of non-zero $E_{err}$, we employ a simplified
Hamiltonian 
\begin{equation}
H\cong\left(\begin{array}{cc}
-E_{err}/2 & -\Delta/2\\
-\Delta/2 & E_{err}/2
\end{array}\right)
\end{equation}
with a discretized 2-dimensional Hilbert space, with the state vector
$\left(\psi_{L},\psi_{R}\right)^{T}$, where $\psi_{L,R}$ denotes
the probability amplitude that the system is in the left or right
potential well. This Hamiltonian is designed to give the right energy
splitting when $E_{err}=0$. The ground and excited energy levels
are $E=\pm\sqrt{\Delta^{2}+E_{err}^{2}}/2$ and corresponding eigenstates
are $\left(\Delta,\mp\sqrt{\Delta^{2}+E_{err}^{2}}-E_{err}\right)^{T}$,
which approximately is $\left(\Delta,\mp\Delta-E_{err}\right)^{T}$
if $E_{err}\ll\Delta$. Further considerations suggest that this type
of inaccuracy manifests itself as an error probability of the order
of $\left(E_{err}/\Delta\right)^{2}$ upon qubit measurement, in EEEM
and charged particle detection applications.

Table AI shows several numerically computed parameters for the rf-SQUID,
with ${C=10.0\:\mathrm{fF}}$, ${L=800\:\mathrm{pH}}$, $\beta=\sqrt{2}\pi/4=1.11$.
Only when estimating $E_{err}$, we assume $\phi_{err}/\phi_{0}=1.0\times10^{-4}$.
To compute the energy splitting $\Delta$, we used a finite step $\delta\theta=\pi/100$
in the Numerov method. We further confirmed that changing the step
size $\delta\theta$ to $\pi/200$ resulted only in negligible changes
of eigenvalues. We used the fact that the ground state and the first
excited state are respectively associated with an even and odd wavefunction.
We were able to determine the eigenvalues by integrating the equation
up to $\theta=2\pi$. The error associated with $\Delta$ corresponds
to $1\%$ changes in $\alpha$ or $\beta$.

\begin{table}
\caption{Computed device parameters for an rf-SQUID with $\beta=1.11$. The
error in the bias flux is assumed to be $1.0\times10^{-4}\phi_{0}$.
The symbol $E_{L}$ stands for $\left(\phi_{0}/2\pi\right)^{2}/2L$.}
\begin{tabular}{|c|c|c|c|c|}
\hline 
$E_{J}/\mathrm{\mu eV}$ & $E_{C}/\mathrm{\mu eV}$ & $E_{L}/\mathrm{\mu eV}$ & $E_{err}/\mathrm{\mu eV}$ & $\Delta/\mathrm{\mu eV}$\tabularnewline
\hline 
\hline 
939 & 32.04 & 423 & 0.83 & 67$\pm$3\tabularnewline
\hline 
\end{tabular}
\end{table}

\section{ELECTROMAGNETICS OF A LONG RECTANGULAR RF-SQUID}

Here we estimate the effective capacitance of a single, long rectangular
rf-SQUID. Figure B1 shows a long rf-SQUID placed along the $x$-axis.
An external, uniform magnetic flux density $\boldsymbol{B}_{ext}$
is generated by a weakly coupled magnet, resulting in magnetic flux
$\phi_{0}/2$ threading the rf-SQUID loop. The long side of the rectangular
rf-SQUID has the length $a$. The center of the rf-SQUID, where the
only Josephson junction (JJ) is located, is at the origin $x=0$.
We treat the rf-SQUID loop as a transmission line. Similar treatments
of superconducting circuit have been described \cite{qbits as spectrometers}.
Let the inductance per unit length be $l$ and likewise capacitance
per unit length be $c$. Let the local current be $i\left(x,t\right)$
and the local linear charge density be $\lambda\left(x,t\right)$.
The sign of these quantities are such that $i\left(x,t\right)>0$
if the current flows towards $+x$ direction on the upper line shown
in Fig. 1, and likewise $\lambda\left(x,t\right)>0$ if positive charge
is present on the upper line. We assume that the charge distribution
on the lower line is the same but with the opposite sign. The uniqueness
of the electric potential implies 
\begin{equation}
\frac{1}{c}\partial_{x}\lambda+l\partial_{t}i=0,\label{eq:gauge invariance}
\end{equation}
when $-a/2<x<0$ or $0<x<a/2$. In the same region, charge conservation
demands $\partial_{t}\lambda+\partial_{x}i=0$. A relation $\lambda\left(\pm a/2,t\right)=0$
holds since the transmission line is short circuited at both the ends.
The origin $x=0$, where the JJ is, requires a separate consideration.
The charge distribution $\lambda\left(x,t\right)$ generally is not
continuous at $x=0$ because a voltage drop can develop across the
JJ. Although Eq. (\ref{eq:gauge invariance}) may seem to imply that
$i\left(x,t\right)$ also is discontinuous, $i\left(x,t\right)$ \emph{is}
continuous because the ``inductance'' due to the JJ is concentrated
to the point $x=0$.

We develop a Lagrangian of the system. We define a field $\phi\left(x,t\right)$
having the dimension of magnetic flux, which is not continuous at
$x=0$, as follows:
\begin{equation}
\phi\left(x,t\right)=\begin{cases}
l\int_{a/2}^{x}dx'i\left(x',t\right) & 0<x\\
l\int_{-a/2}^{x}dx'i\left(x',t\right). & x<0
\end{cases}\label{eq:phi_def}
\end{equation}
We designed the function so that $i=\left(\frac{1}{l}\right)\partial_{x}\phi$
and $\lambda=-c\partial_{t}\phi$ when $-a/2<x<0$ or $0<x<a/2$.
The linear density of electric energy is $\frac{1}{2c}\lambda^{2}=\frac{c}{2}\left(\partial_{t}\phi\right)^{2}$,
which we recognize as kinetic energy of the system because it involves
the time derivative. On the other hand, the linear density of magnetic
energy is $\frac{1}{2l}\left\{ B\left(x,t\right)w-\frac{\phi_{0}}{2a}\right\} ^{2}=\frac{1}{2l}\left(\partial_{x}\phi\right)^{2}$
because of the external field, where $B\left(x,t\right)$ is the magnetic
flux density and $w$ is the width of the transmission line. Hence,
the Lagrangian $L_{TL}$ of the transmission line is
\begin{equation}
L_{TL}=\int_{-a/2}^{a/2}dx\left\{ \frac{c}{2}\left(\partial_{t}\phi\right)^{2}-\frac{1}{2l}\left(\partial_{x}\phi\right)^{2}\right\} ,
\end{equation}
where we skip the point $x=0$ when integrating. The total magnetic
flux generated by $i\left(x,t\right)$ is $\Phi\left(t\right)=\lim_{\varepsilon\rightarrow+0}\left[\phi\left(-\varepsilon,t\right)-\phi\left(\varepsilon,t\right)\right]$.
The Josephson energy stored in the JJ is 
\begin{equation}
-E_{J}\cos\left\{ \frac{\Phi\left(t\right)+\phi_{0}/2}{\left(\phi_{0}/2\pi\right)}\right\} =E_{J}\cos\left(\frac{2\pi\Phi\left(t\right)}{\phi_{0}}\right),
\end{equation}
where $\Phi\left(t\right)+\phi_{0}/2$ is the magnetic flux threading
the rf-SQUID including the externally applied flux. The kinetic energy
associated with the JJ is $C_{J}\dot{\Phi}\left(t\right)^{2}/2$,
where $C_{J}$ is the junction capacitance. The Lagrangian $L_{JJ}$
of the JJ is 
\begin{equation}
L_{JJ}=\frac{C_{J}}{2}\dot{\Phi}\left(t\right)^{2}-E_{J}\cos\left(\frac{2\pi\Phi\left(t\right)}{\phi_{0}}\right).
\end{equation}
 The total Lagrangian we consider is $L_{TL}+L_{JJ}$.

\begin{figure}
\includegraphics[scale=0.2]{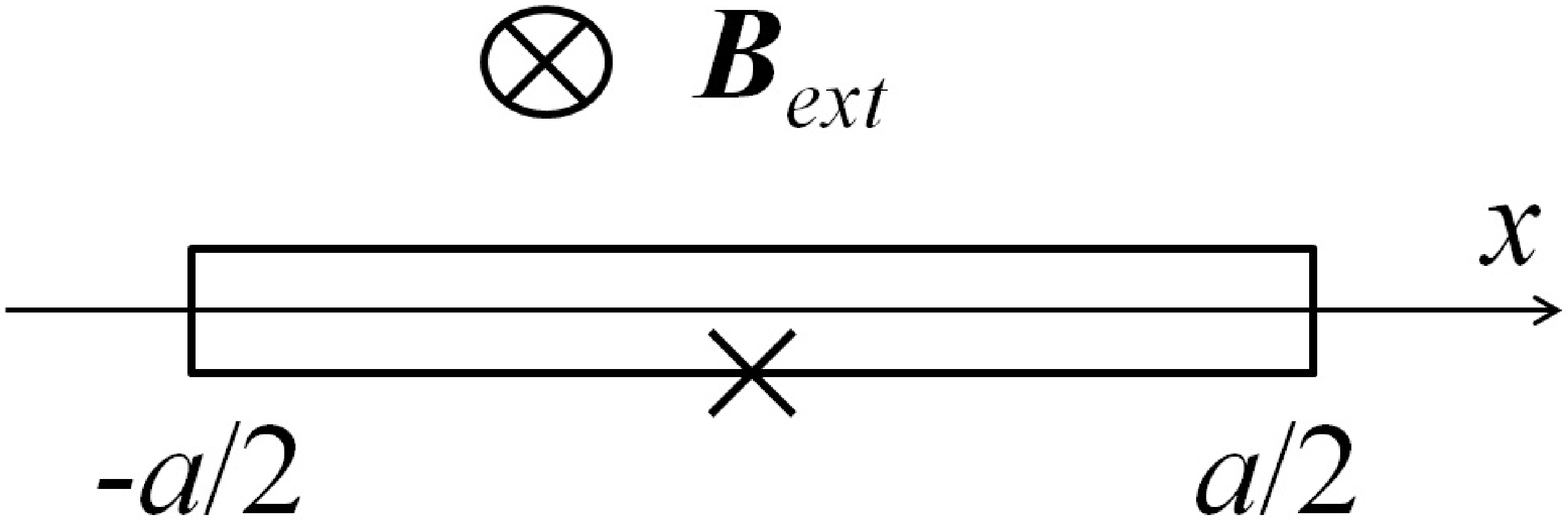}

\caption{Long rectangular rf-SQUID. The ``X'' symbol denotes a Josephson
junction (JJ).}
\end{figure}

To simplify the analysis, we introduce two functions $\psi\left(x,t\right),\xi\left(x,t\right)$
that are antisymmetric and symmetric respectively:
\begin{equation}
\phi\left(-x,t\right)-\phi\left(x,t\right)=\psi\left(x,t\right)\label{eq:psi_def}
\end{equation}
\begin{equation}
\phi\left(-x,t\right)+\phi\left(x,t\right)=\xi\left(x,t\right).
\end{equation}
Since antisymmetric integrand disappears, we obtain 
\begin{equation}
L_{TL}=L_{\psi}+L_{\xi}.\label{eq:separated lagrangian}
\end{equation}
where
\begin{equation}
L_{\psi}=\frac{1}{2}\int_{0}^{a/2}dx\left\{ \frac{c}{2}\left(\partial_{t}\psi\right)^{2}-\frac{1}{2l}\left(\partial_{x}\psi\right)^{2}\right\} 
\end{equation}
and 
\begin{equation}
L_{\xi}=\frac{1}{2}\int_{0}^{a/2}dx\left\{ \frac{c}{2}\left(\partial_{t}\xi\right)^{2}-\frac{1}{2l}\left(\partial_{x}\xi\right)^{2}\right\} .
\end{equation}
Hence, the problem is divided into two non-interacting parts. 

Consider the part involving $\xi\left(x,t\right)$ first. The Euler-Lagrange
equation of motion is the wave equation. The first boundary condition
$\xi\left(a/2,t\right)=0$ comes directly from the definition Eq.
(\ref{eq:phi_def}). On the other hand, we obtain $\partial_{x}\xi\left(x,t\right)=l\left\{ i\left(x,t\right)-i\left(-x,t\right)\right\} $,
which is linear in $x$ for small $x$ because only antisymmetric
part of $i\left(x,t\right)$ contributes to $\partial_{x}\xi\left(x,t\right)$.
Hence the second boundary condition is $\left[\partial_{x}\xi\left(x,t\right)\right]_{x=0}=0$.
The harmonic form of solutions satisfying these boundary conditions
is $\xi\left(x,t\right)\propto\cos\left(n\pi x/a\right)$ for $n=1,3,5,\cdots$.
Quantization of the field $\xi\left(x,t\right)$ results in photons
of various quantized energies, of which the lowest is $\pi\hbar/\left(a\sqrt{lc}\right)$.
The field is free and it does not couple to the JJ under ideal conditions.

Next, we consider the field $\psi\left(x,t\right)$. Note that $\Phi\left(t\right)=\lim_{\varepsilon\rightarrow+0}\psi\left(\varepsilon,t\right)$.
A boundary condition $\psi\left(a/2,t\right)=0$ follows from Eq.
(\ref{eq:phi_def}). The Lagrangian governing this part of the system
is given by $L_{\psi}+L_{JJ}$. To simplify the analysis, we introduce
another field $\hat{\psi}\left(x,t\right)=\psi\left(x,t\right)-\left(1-2x/a\right)\Phi\left(t\right)$,
whose boundary condition is designed to be $\hat{\psi}\left(0,t\right)=\hat{\psi}\left(a/2,t\right)=0$.
In terms of $\hat{\psi}\left(x,t\right)$ and $\Phi\left(t\right)$,
the Lagrangian $L_{\psi}$ is 
\[
L_{\psi}=\frac{1}{2}\int_{0}^{a/2}dx\left\{ \frac{c}{2}\left(\partial_{t}\hat{\psi}\right)^{2}-\frac{1}{2l}\left(\partial_{x}\hat{\psi}\right)^{2}\right\} +\frac{ca}{24}\dot{\Phi}^{2}-\frac{\Phi^{2}}{2la}
\]
\begin{equation}
+\frac{c\dot{\Phi}}{2}\int_{0}^{a/2}dx\left(1-\frac{2x}{a}\right)\left(\partial_{t}\hat{\psi}\right),
\end{equation}
where we used $\int_{0}^{a/2}dx\left(\partial_{x}\hat{\psi}\right)=\left[\hat{\psi}\right]_{x=0}^{x=a/2}=0$.
Combining with $L_{JJ}$, we obtain 
\[
L_{\psi}+L_{JJ}=\frac{C}{2}\dot{\Phi}^{2}-E_{J}\cos\left(\frac{2\pi\Phi\left(t\right)}{\phi_{0}}\right)-\frac{\Phi^{2}}{2L}
\]
\begin{equation}
+\frac{1}{2}\int_{0}^{a/2}dx\mathcal{L},\label{eq:effective lagrangian}
\end{equation}
where we introduced parameters 
\begin{equation}
C=C_{J}+ca/12\label{eq:effective capacitance}
\end{equation}
and $L=la$; and the Lagrangian density $\mathcal{L}$ of the continuous
part of the system is
\begin{equation}
\mathcal{L}=\frac{c}{2}\left(\partial_{t}\hat{\psi}\right)^{2}-\frac{1}{2l}\left(\partial_{x}\hat{\psi}\right)^{2}+c\dot{\Phi}\left(1-\frac{2x}{a}\right)\left(\partial_{t}\hat{\psi}\right).
\end{equation}
The Lagrangian $L_{\psi}+L_{JJ}$ consists of that of a lumped-circuit
rf-SQUID, a free scalar field, and a velocity-dependent coupling between
these. Note that the factor $12$ in Eq. (\ref{eq:effective capacitance})
is a consequence of our particular choice of $\hat{\psi}\left(x,t\right)$,
although the choice does simplify the form of the Lagrangian (\ref{eq:effective lagrangian}).

Full analysis of the system described in Eq. (\ref{eq:effective lagrangian})
would be complex and is beyond the scope of the present work. However,
few preliminary remarks are in order. Pretend for the moment that
the coupling term $c\dot{\Phi}\left(1-\frac{2x}{a}\right)\left(\partial_{t}\hat{\psi}\right)$
is absent. The field $\hat{\psi}$ has the harmonic form of solutions
$\hat{\psi}\propto\sin\left(2n\pi x/a\right)$, where $n$ is a positive
integer. This means that the minimum excitation energy of a photon
is 
\begin{equation}
\frac{2\pi\hbar}{a\sqrt{lc}}=\frac{4.1\:\mathrm{meV}}{\sqrt{\left(la/\mathrm{nH}\right)\left(ca/\mathrm{fF}\right)}}.\label{eq:photon energy}
\end{equation}
Hence in a typical parameter range, especially when the length $a$
of the rf-SQUID is not too large, the photon energy is higher than
the qubit energy scale $\Delta$. This makes it likely that the photon
degrees of freedom stay in the vacuum state adiabatically during the
slow motion of the JJ degrees of freedom, not unlike cases in which
the Born-Oppenheimer approximation is valid. A related phenomenon
has been reported in the context of macroscopic quantum tunneling
\cite{MQT latency}.

The capacitance per unit length between the upper and the lower line
of the circuit may be estimated by using the formula for two parallel
thin rods 
\begin{equation}
c=\frac{\pi\varepsilon_{eff}}{\ln\left(d/a\right)},
\end{equation}
where $\varepsilon_{eff}$ is the effective electric permittivity,
$d\cong10\mu\mathrm{m}$ is the distance between the rods, and $a$
is the radius of the rods that should satisfy $a\ll d$. If the qubit
is fabricated on a silicon wafer (possibly with a thin $\mathrm{Si}\mathrm{O}_{2}$
layer) then we obtain $\varepsilon_{eff}\cong\left(\varepsilon_{0}+\varepsilon_{\mathrm{Si}}\right)/2=6.34\varepsilon_{0}$.
Ignoring the logarithmic factor, we obtain $c/12\cong{14.7\:\mathrm{fF}/\mathrm{mm}}$.
Note that the factor $12$ appeared in Eq. (\ref{eq:effective capacitance}).

Next, we consider a more favorable scenario. The value of $\varepsilon_{eff}$
could be significantly lowered by etching the Si wafer between the
lines, possibly by deep reactive ion etching (DRIE) with the depth
comparable to, or more than, $d$. (This could be beneficial also
from the perspective of charged particle optical coherence. At least
at the room temperature, decoherence of electron waves is observed
when the electrons travel the distance of ${10\:\mathrm{mm}}$ along
a semiconducting surface, while keeping the distance of several $\mu\mathrm{m}$
from the surface \cite{Hasselbach}.) Furthermore, one could make
an effort to make the factor $\ln\left(d/a\right)$ comparable to
$\pi$. Provided that these measures can be made, then the capacitance
per unit length is given simply as $c'\cong\varepsilon_{0}$. We obtain
$c'/12\cong{0.74\:\mathrm{fF}/\mathrm{mm}}$ in this case.

\section{POTENTIAL ENERGY OF A SET OF WEAKLY-COUPLED RF-SQUIDS}

We first make a minor digression and exhibit somewhat elementary results
about a system of inductors. This is for the convenience of the reader,
and also for showing approximations that we use. Let $A$ be a square
matrix and $\boldsymbol{b},\boldsymbol{c}$ be vectors. It is straightforward
to verify that 
\begin{equation}
\left[\begin{array}{cc}
A & \boldsymbol{b}\\
\boldsymbol{c}^{T} & d
\end{array}\right]^{-1}=\left[\begin{array}{cc}
A^{-1}+\Delta^{-1}A^{-1}\boldsymbol{b}\boldsymbol{c}^{T}A^{-1} & -\Delta^{-1}A^{-1}\boldsymbol{b}\\
-\Delta^{-1}\boldsymbol{c}^{T}A^{-1} & \Delta^{-1}
\end{array}\right],\label{eq:matrix inversion}
\end{equation}
where the Schur complement is given as $\Delta=d-\boldsymbol{c}^{T}A^{-1}\boldsymbol{b}$.
Now consider a system of coupled ``internal'' inductors, with the
inductance matrix $l$ and the associated current vector $\boldsymbol{i}$
and the magnetic flux vector $\boldsymbol{\phi}$. This entire system
is weakly coupled to a large ``external'' inductor $L$, with the
current $I$ and the magnetic flux $\Phi$, that acts as a bias magnetic
flux generator. The enlarged inductance matrix satisfies 
\begin{equation}
\left[\begin{array}{c}
\boldsymbol{\phi}\\
\Phi
\end{array}\right]=\left[\begin{array}{cc}
l & \boldsymbol{m}\\
\boldsymbol{m}^{T} & L
\end{array}\right]\left[\begin{array}{c}
\boldsymbol{i}\\
I
\end{array}\right],
\end{equation}
where values of mutual inductance are represented in the vector $\boldsymbol{m}$.
The magnetic energy of the system is given as \begin{widetext} 
\begin{equation}
U_{m}=\frac{1}{2}\left[\begin{array}{cc}
\boldsymbol{i}^{T} & I\end{array}\right]\left[\begin{array}{cc}
l & \boldsymbol{m}\\
\boldsymbol{m}^{T} & L
\end{array}\right]\left[\begin{array}{c}
\boldsymbol{i}\\
I
\end{array}\right]=\frac{1}{2}\left[\begin{array}{cc}
\boldsymbol{\phi}^{T} & \Phi\end{array}\right]\left[\begin{array}{cc}
l & \boldsymbol{m}\\
\boldsymbol{m}^{T} & L
\end{array}\right]^{-1}\left[\begin{array}{c}
\boldsymbol{\phi}\\
\Phi
\end{array}\right].
\end{equation}
When applying the matrix inversion formula Eq. (\ref{eq:matrix inversion}),
we use the fact that the bias magnet is weakly coupled, and hence
ignore the second and higher order terms with respect to the mutual
inductance $\boldsymbol{m}$. Hence we obtain 
\begin{equation}
U_{m}\cong\frac{1}{2}\left[\begin{array}{cc}
\boldsymbol{\phi}^{T} & \Phi\end{array}\right]\left[\begin{array}{cc}
l^{-1} & -l^{-1}\boldsymbol{m}/L\\
-\boldsymbol{m}^{T}l^{-1}/L & 1/L
\end{array}\right]\left[\begin{array}{c}
\boldsymbol{\phi}\\
\Phi
\end{array}\right]=\frac{1}{2}\boldsymbol{\phi}^{T}l^{-1}\boldsymbol{\phi}-\frac{\Phi}{L}\boldsymbol{m}^{T}l^{-1}\boldsymbol{\phi}+\frac{\Phi^{2}}{2L}.
\end{equation}
 \end{widetext} Since we decided to ignore higher order terms in
$\boldsymbol{m}$, this expression is further approximated as 
\begin{equation}
U_{m}\cong\frac{1}{2}\left(\boldsymbol{\phi}-\frac{\Phi}{L}\boldsymbol{m}\right)^{T}l^{-1}\left(\boldsymbol{\phi}-\frac{\Phi}{L}\boldsymbol{m}\right)+\frac{\Phi^{2}}{2L},\label{eq:magnetic energy matrix form}
\end{equation}
where the second term will be neglected in the followings. (Such an
omission may not always be justified. See Ref. \cite{Ralph et al}.)
Clearly, $\frac{\Phi}{L}\boldsymbol{m}\cong\boldsymbol{m}I$ represents
the external bias flux and we regard $\boldsymbol{\phi}-\frac{\Phi}{L}\boldsymbol{m}$
as a set of magnetic flux generated by the current flowing in each
internal inductor.

Returning to our system, a magnetic flux $\phi_{k}+\phi_{0}/2$ threads
the $k-$th rf-SQUID. The first term $\phi_{k}$ is generated by the
current in the $k-$th rf-SQUID ring, while the second term $\phi_{0}/2$
is the bias flux applied externally. We find (See Eq. (\ref{eq:magnetic energy matrix form}))
the magnetic energy stored in the system, if we can invert the inductance
matrix of the system of coupled rf-SQUIDs. The inductance matrix may
be written in the form of $L\mathbf{A}$, where the square tridiagonal
matrix $\mathbf{A}$ with elements $a_{i,k}$ has the diagonal entries
$a_{i,i}=1$ and the adjacent off-diagonal entries $a_{i,i+1}=a_{i,i-1}=M/L$.
In order to invert this, we make an assumption $M/L\ll1$. To the
first order in $\delta$, $\mathbf{B}=\mathbf{A}^{-1}$ has the diagonal
entries $b_{i,i}=1$ and the off-diagonal entries $b_{i,i+1}=b_{i,i-1}=-M/L$,
while all other entries are zero. Hence
\[
U_{m}\cong\frac{1}{2L}\sum_{k=1}^{N}\phi_{k}^{2}-\frac{M}{L^{2}}\sum_{k=1}^{N-1}\phi_{k}\phi_{k+1}
\]

\begin{equation}
\cong\frac{1}{2\left(L+2M\right)}\sum_{k=1}^{N}\phi_{k}^{2}+\frac{M}{2L^{2}}\sum_{k=1}^{N-1}\left(\phi_{k+1}-\phi_{k}\right)^{2},
\end{equation}
to the first order in $M/L$. We may argue that the level of accuracy
is maintained by this ``approximation'' anyway, because non-zero
mutual inductance beyond the nearest neighbor rf-SQUIDs should exist
in the first place.

The condition $M/L\ll1$ is satisfied in the parameter region of interest.
The rf-SQUID mentioned in the main text has device parameters $C={10.0\:\mathrm{fF}}$,
${L=800\:\mathrm{pH}}$ and $\beta=1.11$, for which $\Delta={1.1\times10^{-23}\:\mathrm{J}}$.
On the other hand, we expect to use the composite qubit in the region
$R\cong1$ near the QCP, which translates to $J=M\phi_{0}^{2}/2N^{2}L^{2}\cong\Delta/2$,
or equivalently $M/L\cong LN^{2}\Delta/\phi_{0}^{2}$. The right hand
side turns out to be $3.2\times10^{-2}$, which justifies the use
of the aforementioned condition $M/L\ll1$. (This argument is not
circular as one could have picked such $M/L$ first, then developed
approximate expressions based on it, and later found that $R$ happens
to be close to $1$.)

\section{RENORMALIZING THE TRANSVERSE-FIELD ISING MODEL WITH A SMALL LONGITUDINAL
FIELD}

We extend the renormalization analysis of the transverse-field Ising
model with a zero longitudinal-field (i.e. $\varepsilon=0$), originally
due to Fernandez-Pacheco \cite{Fernandez-Pacheco,Monthus exposition},
to the $\varepsilon\neq0$ case. The straightforward but lengthy process
of extension is presented below. Note that detailed studies on the
transverse-field Ising system with a longitudinal field have appeared
in the literature \cite{longitudinal field and trans-field Ising}.

For block renormalization purposes, we rewrite the Hamiltonian (Eq.
(2) in the main text) 
\begin{equation}
H=-J\sum_{i}\sigma_{i}^{z}\sigma_{i+1}^{z}-h\sum_{i}\sigma_{i}^{x}-\varepsilon\sum_{i}\sigma_{i}^{z},
\end{equation}
where the index $i$ points to each spin, as a sum of intra-block
and inter-block terms 
\begin{equation}
H=\sum_{j}H_{j}^{intra}+\sum_{j}H_{j}^{inter},
\end{equation}
\begin{equation}
H_{j}^{intra}=-h\sigma_{2j-1}^{x}-\varepsilon\sigma_{2j-1}^{z}-J\sigma_{2j-1}^{z}\sigma_{2j}^{z},\label{eq:intra hamiltonian}
\end{equation}
\begin{equation}
H_{j}^{inter}=-h\sigma_{2j}^{x}-\varepsilon\sigma_{2j}^{z}-J\sigma_{2j-2}^{z}\sigma_{2j-1}^{z},
\end{equation}
where $j$ points to each block of $2$ spins. Notations such as $\sigma_{i}^{z}$
imply identity operators for all spins except the spin $i$. We call
spins with an odd index ``slave spins'' and the rest ``master spins''.
The Hilbert space pertaining to the $i$-th spin is spanned by $|s\rangle_{i}$,
where $s=\pm1$ denotes one of the eigenvalues of the operator $\sigma_{i}^{z}$.
The identity operator pertaining to the $i$-th spin is denoted $I_{i}$.

We first focus on the intra-block Hamiltonian (\ref{eq:intra hamiltonian})
and seek energy eigenstates of the form $\left(a|1\rangle_{2j-1}+b|-1\rangle_{2j-1}\right)\otimes|s\rangle_{2j}$.
We find eigenvalues 
\begin{equation}
\lambda_{\pm,s}=\pm\sqrt{h^{2}+\left(sJ+\varepsilon\right)^{2}},
\end{equation}
and respectively corresponding eigenvectors $|\lambda_{\pm,s}\rangle_{2j-1}\otimes|s\rangle_{2j}$,
where \begin{widetext}
\begin{equation}
|\lambda_{\pm,s}\rangle_{2j-1}=F\left\{ -h|1\rangle_{2j-1}+\left[\left(sJ+\varepsilon\right)\pm\sqrt{h^{2}+\left(sJ+\varepsilon\right)^{2}}\right]|-1\rangle_{2j-1}\right\} ,
\end{equation}
where $F$ is a real and positive normalization factor that depends
on $s$ and $\lambda_{\pm}$. An equivalent expression is 
\begin{equation}
|\lambda_{\pm,s}\rangle_{2j-1}=F\left\{ \left[\left(sJ+\varepsilon\right)\mp\sqrt{h^{2}+\left(sJ+\varepsilon\right)^{2}}\right]|1\rangle_{2j-1}+h|-1\rangle_{2j-1}\right\} ,
\end{equation}
\end{widetext}which may be useful in the $h\rightarrow0$ limit for
some combinations of $s$ and $\lambda_{\pm}$.

We block-renormalize the system by demanding that each slave spin
$2j-1$ is in the ground state $|\lambda_{-,s}\rangle_{2j-1}$of the
associated intra-block Hamiltonian $H_{j}^{intra}$. To avoid cluttered
presentation, we use $|g_{s}\rangle_{2j-1}\equiv|\lambda_{-,s}\rangle_{2j-1}$
and $g_{s}\equiv\lambda_{-,s}$ below. Note that the state $|g_{s}\rangle_{2j-1}$
depends on the state $|s\rangle_{2j}$ of the associated master spin
$2j$. To be concrete, the renormalized Hamiltonian $H_{R}$ is defined
as 
\begin{equation}
H_{R}=P^{\dagger}HP,
\end{equation}
where $P=\otimes_{j}P_{j}$ and 
\begin{equation}
P_{j}=|g_{1}\rangle_{2j-1}\otimes\left(|1\rangle_{2j}\langle1|_{2j}\right)+|g_{-1}\rangle_{2j-1}\otimes\left(|-1\rangle_{2j}\langle-1|_{2j}\right),
\end{equation}
although one might prefer to use states of $j$-th renormalized spin
$\langle\pm1_{R}|_{j}$ instead of $\langle\pm1|_{2j}$.

We use the following relations to obtain $H_{R}$. \begin{widetext}

\begin{equation}
P_{j}^{\dagger}H_{j}^{intra}P_{j}=g_{1}|1\rangle_{2j}\langle1|_{2j}+g_{-1}|-1\rangle_{2j}\langle-1|_{2j},
\end{equation}
\begin{equation}
P_{j}^{\dagger}\sigma_{2j}^{z}P_{j}=P_{j}^{\dagger}\left(|1\rangle_{2j}\langle1|_{2j}-|-1\rangle_{2j}\langle-1|_{2j}\right)P_{j}=|1\rangle_{2j}\langle1|_{2j}-|-1\rangle_{2j}\langle-1|_{2j}=\sigma_{2j}^{z},
\end{equation}
\begin{equation}
P_{j}^{\dagger}\sigma_{2j}^{x}P_{j}=\langle g_{-1}|g_{1}\rangle_{2j-1}|-1\rangle_{2j}\langle1|_{2j}+\langle g_{1}|g_{-1}\rangle_{2j-1}|1\rangle_{2j}\langle-1|_{2j}=\alpha\sigma_{2j}^{x},
\end{equation}
\end{widetext}where 
\begin{equation}
\alpha=\langle g_{-1}|g_{1}\rangle_{2j-1}=\langle g_{1}|g_{-1}\rangle_{2j-1}
\end{equation}
and 
\[
P_{j}^{\dagger}P_{j-1}^{\dagger}\sigma_{2j-2}^{z}\sigma_{2j-1}^{z}P_{j-1}P_{j}
\]
\begin{equation}
=\left(P_{j-1}^{\dagger}\sigma_{2j-2}^{z}P_{j-1}\right)\left(P_{j}^{\dagger}\sigma_{2j-1}^{z}P_{j}\right),
\end{equation}
\begin{equation}
P_{j}^{\dagger}\sigma_{2j-1}^{z}P_{j}=\beta_{1}|1\rangle_{2j}\langle1|_{2j}+\beta_{-1}|-1\rangle_{2j}\langle-1|_{2j},
\end{equation}
where 
\begin{equation}
\beta_{1}=\langle g_{1}|\sigma_{2j-1}^{z}|g_{1}\rangle_{2j-1},\:\beta_{-1}=\langle g_{-1}|\sigma_{2j-1}^{z}|g_{-1}\rangle_{2j-1}.
\end{equation}
To proceed further, we assume $\varepsilon\ll h,J$ and retain only
the first order terms in $\varepsilon$ in the following expressions.
We obtain 
\begin{equation}
g_{\pm1}=-\sqrt{h^{2}+J^{2}}\left(1\pm\frac{J\varepsilon}{h^{2}+J^{2}}\right),
\end{equation}
 \begin{widetext}

\begin{equation}
|g_{1}\rangle=F_{1}\left\{ -h|1\rangle_{2j-1}+\left[J-\sqrt{h^{2}+J^{2}}+\varepsilon\left(1-\frac{J}{\sqrt{h^{2}+J^{2}}}\right)\right]|-1\rangle_{2j-1}\right\} ,
\end{equation}

\begin{equation}
|g_{-1}\rangle=F_{-1}\left\{ -h|1\rangle_{2j-1}-\left[J+\sqrt{h^{2}+J^{2}}-\varepsilon\left(1+\frac{J}{\sqrt{h^{2}+J^{2}}}\right)\right]|-1\rangle_{2j-1}\right\} ,
\end{equation}
where 
\begin{equation}
F_{\pm1}=\frac{1}{\sqrt{h^{2}+\left(J\mp\sqrt{h^{2}+J^{2}}\right)^{2}}}\left\{ 1+\frac{\varepsilon}{\sqrt{h^{2}+J^{2}}}\frac{\left(J\mp\sqrt{h^{2}+J^{2}}\right)^{2}}{h^{2}+\left(J\mp\sqrt{h^{2}+J^{2}}\right)^{2}}\right\} ,
\end{equation}
\end{widetext}and 
\begin{equation}
F_{1}F_{-1}=\frac{1}{2h\sqrt{h^{2}+J^{2}}}\left(1+\frac{\varepsilon}{\sqrt{h^{2}+J^{2}}}\right).
\end{equation}
We further obtain

\begin{equation}
\alpha=F_{1}F_{-1}2h^{2}\left(1-\frac{\varepsilon}{\sqrt{h^{2}+J^{2}}}\right)=\frac{h}{\sqrt{h^{2}+J^{2}}}.
\end{equation}
This is a property along the $x$ axis, which should not be affected
by the parameter $\varepsilon$ linearly, because nonzero $\varepsilon$
breaks symmetry with respect to the $z$ axis. Continuing, using an
identity 
\begin{equation}
\frac{J^{2}\pm J\sqrt{h^{2}+J^{2}}}{h^{2}+J^{2}\pm J\sqrt{h^{2}+J^{2}}}=\pm\frac{J}{\sqrt{h^{2}+J^{2}}},
\end{equation}
we obtain \begin{widetext}
\[
\beta_{\pm1}=F_{\pm1}^{2}\left\{ h^{2}-\left(J\mp\sqrt{h^{2}+J^{2}}\right)^{2}+\frac{2\varepsilon\left(J\mp\sqrt{h^{2}+J^{2}}\right)^{2}}{\sqrt{h^{2}+J^{2}}}\right\} 
\]
\end{widetext}
\[
=\pm\frac{J}{\sqrt{h^{2}+J^{2}}}\left\{ 1+\frac{2\varepsilon}{\sqrt{h^{2}+J^{2}}}\frac{2h^{2}\left(J\mp\sqrt{h^{2}+J^{2}}\right)^{2}}{h^{4}-\left(J\mp\sqrt{h^{2}+J^{2}}\right)^{4}}\right\} 
\]
\begin{equation}
=\pm\frac{J}{\sqrt{h^{2}+J^{2}}}\left\{ 1\pm\frac{\varepsilon h^{2}}{J\left(h^{2}+J^{2}\right)}\right\} .
\end{equation}
Combining these results, we obtain 
\begin{equation}
P_{j}^{\dagger}H_{j}^{intra}P_{j}=-\sqrt{h^{2}+J^{2}}I_{2j}-\frac{J\varepsilon}{\sqrt{h^{2}+J^{2}}}\sigma_{2j}^{z},
\end{equation}
where the first term is an unimportant additive constant and we will
omit it below. We further get 
\begin{equation}
P_{j}^{\dagger}\sigma_{2j-1}^{z}P_{j}=\frac{\varepsilon h^{2}}{\left(h^{2}+J^{2}\right)^{3/2}}I_{2j}+\frac{J}{\sqrt{h^{2}+J^{2}}}\sigma_{2j}^{z}
\end{equation}
and hence \begin{widetext}
\[
P_{j}^{\dagger}P_{j-1}^{\dagger}H_{j}^{inter}P_{j-1}P_{j}=-P_{j}^{\dagger}P_{j-1}^{\dagger}\left(h\sigma_{2j}^{x}+\varepsilon\sigma_{2j}^{z}+J\sigma_{2j-2}^{z}\sigma_{2j-1}^{z}\right)P_{j-1}P_{j}
\]
\begin{equation}
=-\left(\frac{h^{2}}{\sqrt{h^{2}+J^{2}}}\sigma_{2j}^{x}+\varepsilon\sigma_{2j}^{z}+\frac{J^{2}}{\sqrt{h^{2}+J^{2}}}\sigma_{2j-2}^{z}\sigma_{2j}^{z}+\frac{\varepsilon Jh^{2}}{\left(h^{2}+J^{2}\right)^{3/2}}\sigma_{2j-2}^{z}\right).
\end{equation}
\end{widetext}Thus, the renormalized Hamiltonian has the same form
as the original one: 
\begin{equation}
H_{R}=-J_{R}\sum_{j}\sigma_{2j}^{z}\sigma_{2j+2}^{z}-h_{R}\sum_{j}\sigma_{2j}^{x}-\varepsilon_{R}\sum_{j}\sigma_{2j}^{z},
\end{equation}
where renormalized parameters are 
\begin{equation}
J_{R}=\frac{J^{2}}{\sqrt{h^{2}+J^{2}}},
\end{equation}
\begin{equation}
h_{R}=\frac{h^{2}}{\sqrt{h^{2}+J^{2}}},
\end{equation}
\begin{equation}
\varepsilon_{R}=\varepsilon\left\{ 1+\frac{J}{\sqrt{h^{2}+J^{2}}}\left(1+\frac{h^{2}}{h^{2}+J^{2}}\right)\right\} .
\end{equation}

\section{EXPECTED ERRORS IN ENTANGLEMENT-ENHANCED ELECTRON MICROSCOPY: A PRELIMINARY
ANALYSIS}

In EEEM, the composite qubit ideally holds magnetic flux of either
zero or $\phi_{0}$. As mentioned in the main text, the magnetic flux
difference between the two qubit states may be slightly different
from the ideal $\phi_{0}$. This is the first kind of error, which
could be nullified by adjusting $\beta$ of each constituent qubit.
However, we should like to see how sensitive the system is to such
an error. The second kind of nonideality is that the magnetic flux
held by the composite qubit may not be quantum mechanically well-defined.
Below we study these two kinds of errors in turn. However, the problem
is complex and our study presented here is preliminary.

To evaluate the first kind of error, we assume that the qubit holds
magnetic flux of either $\left(1-P\right)\phi_{0}/2$ or $\left(1+P\right)\phi_{0}/2$,
where the real parameter $P$ is close to, but smaller than, the ideal
value $1$. Note that the difference is $P\phi_{0}$. We adjusted
their mean to be $\phi_{0}/2$ for later convenience. This adjustment
corresponds to shifting of the phase of the electron waves going through
the magnetic flux ring, which most likely means simply a shift of
the focus of the objective lens in practice. Let $\zeta$ be $\left(\pi/4\right)\left(1-P\right)$,
which is small. Following the EEEM literature, the two qubit states
with flux $\left(1-P\right)\phi_{0}/2$ and $\left(1+P\right)\phi_{0}/2$
are respectively denoted by $|0\rangle$ and $|1\rangle$, although
these are denoted by $\mid\Uparrow'\rangle$ and $\mid\Downarrow'\rangle$
in the main text. Due to the AB effect, with the qubit state $|1\rangle$
the electron passing through the magnetic flux ring (the electron
state $|a\rangle_{e}=\left(|0\rangle_{e}-|1\rangle_{e}\right)\sqrt{2}$)
receives a phase factor $-e^{-i\zeta}$ whereas the electron passing
by the ring (the electron state $|s\rangle_{e}=\left(|0\rangle_{e}+|1\rangle_{e}\right)\sqrt{2}$)
receives a phase factor $e^{i\zeta}$. To make analysis simpler, we
assume that a small amount of vector potential $\boldsymbol{A}$ (in
the Coulomb gauge. See arguments in Ref. \cite{q-interface}) remains
$outside$ the magnetic flux ring generated by the qubit, in the following
way: When the qubit state is $|0\rangle$, then the electron states
$|a\rangle_{e}$ and $|s\rangle_{e}$ receives phase factors $e^{i\zeta}$
and $e^{-i\zeta}$, respectively. Put differently, the difference
in the phase shifts, for electrons in the states $|a\rangle_{e}$
and $|s\rangle_{e}$, is less than the ideal $\pi$ in the case of
the qubit state $|1\rangle$, but the mean phase shift is assumed
to be $\pi/2$; whereas the phase shift difference is more than the
ideal $0$ when the qubit is in the state $|0\rangle$ but the mean
phase shift is assumed to be $0$. The following formulae should be
modified if this simplifying assumption is not used.

The initial states for the electron and the qubit are respectively
$|0\rangle_{e}$ and $|s\rangle=\left(|0\rangle+|1\rangle\right)/\sqrt{2}$.
After interaction, the state of the composite system becomes 
\[
\cos\zeta\frac{|00\rangle+|11\rangle}{\sqrt{2}}+i\sin\zeta\frac{|01\rangle-|10\rangle}{\sqrt{2}}
\]
\begin{equation}
\cong\frac{|00\rangle+|11\rangle}{\sqrt{2}}+i\zeta\frac{|01\rangle-|10\rangle}{\sqrt{2}},
\end{equation}
where $|00\rangle=|0\rangle_{e}\otimes|0\rangle$ etc. Then the electron
wave goes through the specimen and the state $|0\rangle_{e}$ receives
a phase factor $e^{2is}$ relative to the state $|1\rangle_{e}$.
Hence we have 
\begin{equation}
\frac{e^{is}|00\rangle+e^{-is}|11\rangle}{\sqrt{2}}+i\zeta\frac{e^{is}|01\rangle-e^{-is}|10\rangle}{\sqrt{2}}.
\end{equation}
For simplicity, we assume that the electron is detected either in
the state $|s\rangle_{e}$ or $|a\rangle_{e}$; or in other words
$\left(|0\rangle_{e}\pm|1\rangle_{e}\right)/\sqrt{2}$. The qubit
is left in respective states 
\[
\frac{e^{is}|0\rangle\pm e^{-is}|1\rangle}{\sqrt{2}}+i\zeta\frac{e^{is}|1\rangle\mp e^{-is}|0\rangle}{\sqrt{2}}
\]
\begin{equation}
=\frac{1}{\sqrt{2}}\left\{ \left(e^{is}\mp i\zeta e^{-is}\right)|0\rangle\pm\left(e^{-is}\pm i\zeta e^{is}\right)|1\rangle\right\} .\label{eq:messed up qubit}
\end{equation}
The effect of non-zero $\zeta$ must be corrected unless $\zeta\ll s$.
Since $s$ is not known \emph{a priori}, one may need to employ approximations
such as $i\zeta e^{-is}\cong i\zeta$ on grounds that both the parameters
$\zeta$ and $s$ are small. Further study is needed to determine
the region of validity of such approximations. Note that, in high
resolution cryoelectron microscopy of biological specimens, we deal
with values of $s$ as small as $0.01$ \cite{adaptive EM}. Further
note that factors such as $\left(e^{-is}\pm i\zeta e^{is}\right)$
are multiplied a few tens of times in EEEM before the qubit is measured,
although we only discussed the case of single factors for the sake
of simplicity.

We turn to the second issue of the qubit states that do not have definite
magnetic flux values, but have quantum mechanically smeared values.
We would need to analyze the composite qubit not far from the QCP
to obtain the details of such states for a detailed study. Instead,
in order to see the general structure of the problem, here we schematically
write the state corresponding to Eq. (\ref{eq:messed up qubit}) as
\[
\left\{ c\left(e^{is}\mp i\zeta e^{-is}\right)|0\rangle+c'\left(e^{is}\mp i\zeta'e^{-is}\right)|0'\rangle+\cdots\right\} 
\]
\begin{equation}
\pm\left\{ d\left(e^{-is}\pm i\zeta e^{is}\right)|1\rangle+d'\left(e^{-is}\pm i\zeta'e^{is}\right)|1'\rangle+\cdots\right\} ,\label{eq:messed up qubit 2}
\end{equation}
where the primes indicate entities with slightly different magnetic
flux values with respect to the entities without the primes; and $c,d$
are quantum amplitudes. On the other hand, if the qubit's basis states
are written as 
\begin{equation}
|\mathbf{0}\rangle=c|0\rangle+c'|0'\rangle+\cdots,\:|\mathbf{1}\rangle=d|1\rangle+d'|1'\rangle+\cdots,\label{eq:qubit base state}
\end{equation}
then the state in Eq. (\ref{eq:messed up qubit 2}) is written as,
up to a normalization factor 
\begin{equation}
e^{is+\zeta_{0}}|\mathbf{0}\rangle+e^{-is+\zeta_{1}}|\mathbf{1}\rangle+\sum|\textrm{excited states}\rangle.
\end{equation}
Further study is required to find the phase shifts $\zeta_{0},\zeta_{1}$
that must be corrected, as well as the probability of the qubit to
get excited, which causes an error that is unlikely to be correctable.
Note that values of $\zeta_{0},\zeta_{1}$ depend on the outcomes
of electron detection events and also weakly on the unknown parameter
$s$.

\section{EVOLUTION OF HAMILTONIAN PARAMETERS UPON RENORMALIZATION}

The Hamiltonian parameters $h$ and $\varepsilon$ change to renormalized
values $h''$ and $\varepsilon''$ after $2$ iterations of block
renormalization. Figure F1 shows ratios $h''/h$ and $\varepsilon''/\varepsilon$
plotted against the parameter $R$ of the original system. The plotted
functions are 
\begin{equation}
\frac{h''}{h}=\frac{R}{\sqrt{1+R^{2}}}\cdot\frac{R^{2}}{\sqrt{1+R^{4}}},
\end{equation}
and
\begin{equation}
\frac{\varepsilon''}{\varepsilon}=\left(1+\frac{1+2R^{2}}{\left(1+R^{2}\right)^{3/2}}\right)\left(1+\frac{1+2R^{4}}{\left(1+R^{4}\right)^{3/2}}\right).
\end{equation}
The final renormalized spin corresponds to $N=2^{2}=4$ spins in the
original system.\setcounter{figure}{0} 
\begin{figure}[t]
\includegraphics[scale=0.3]{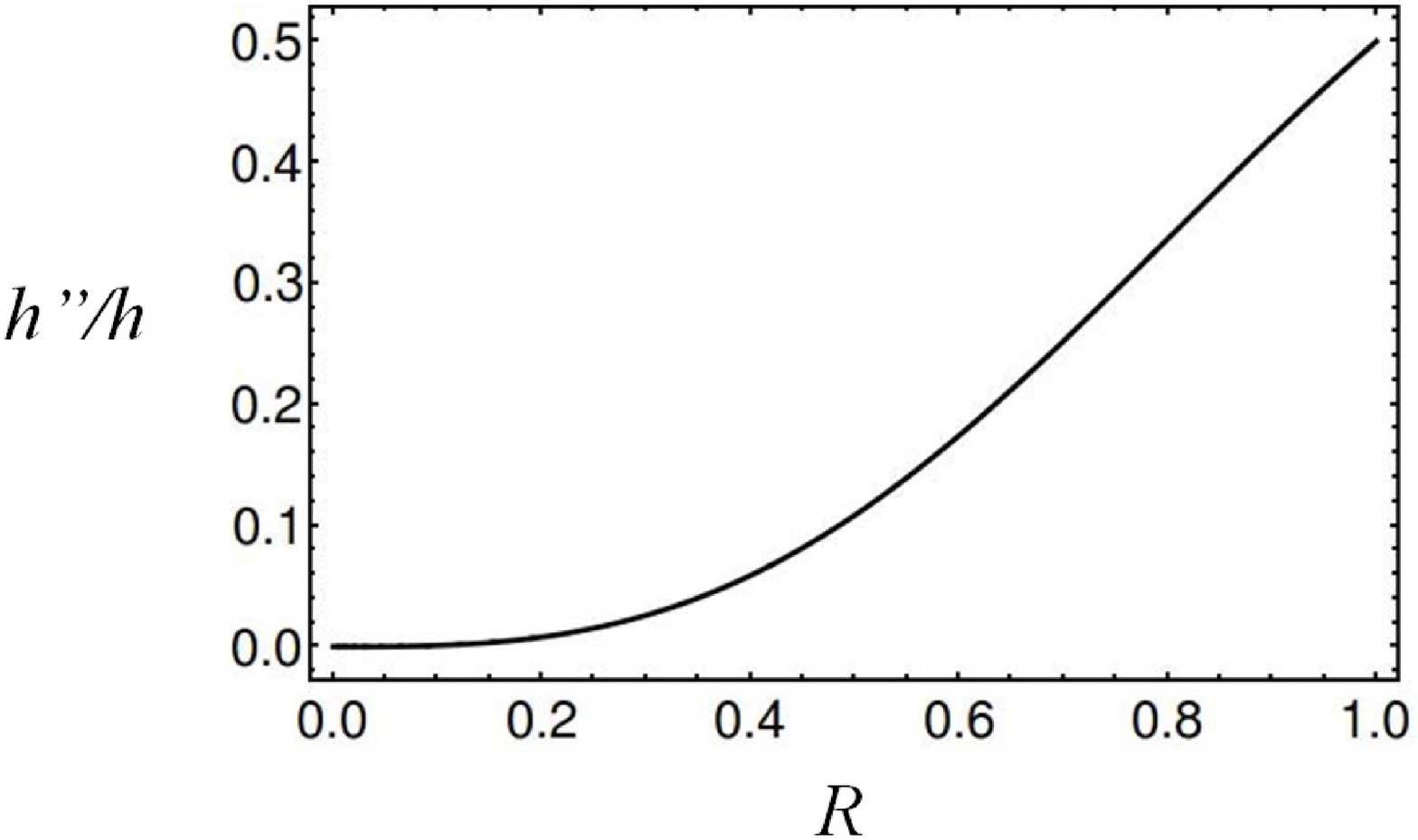} (a)

\includegraphics[scale=0.3]{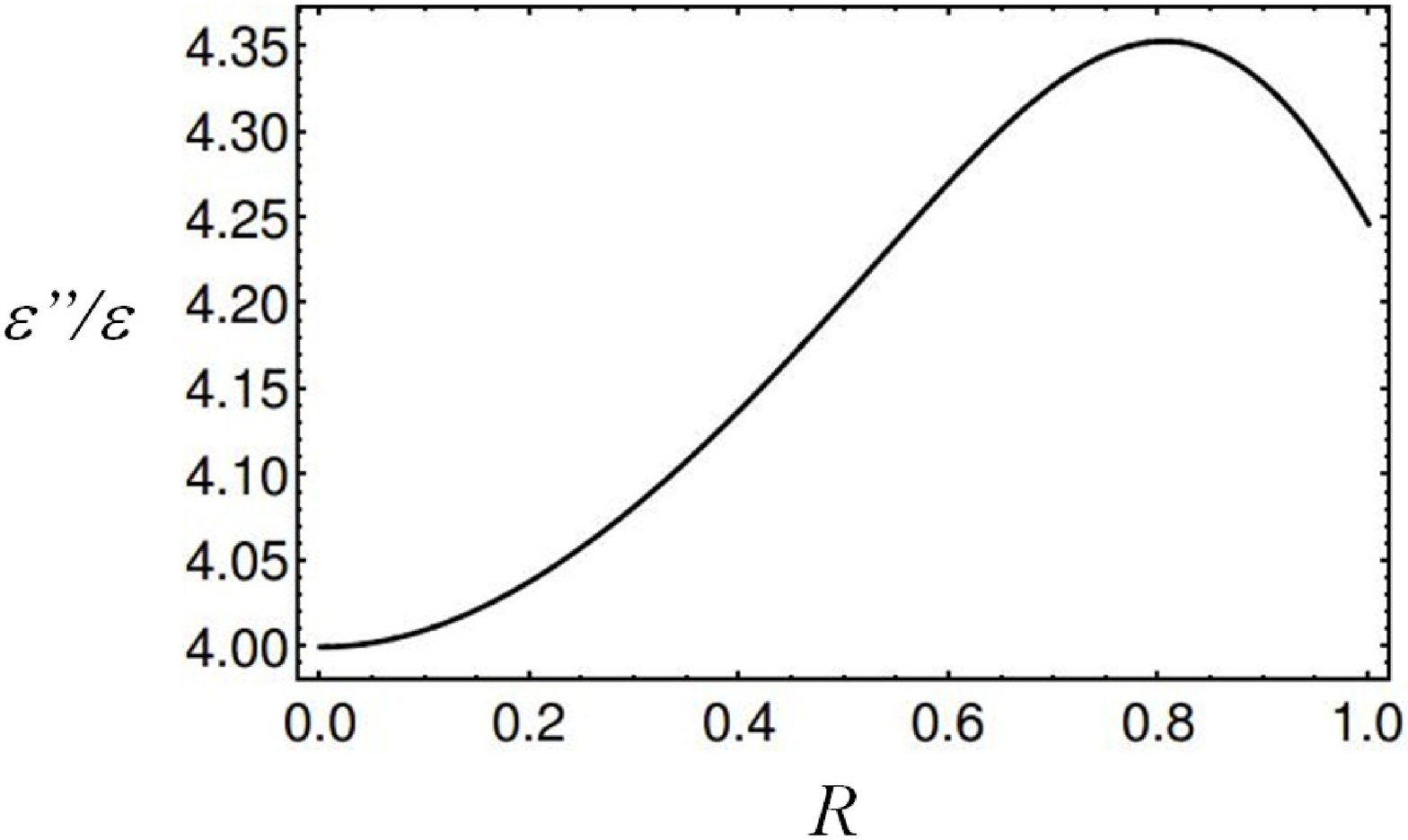} (b)

\caption{Changes of the parameters $h$ and $\varepsilon$ after $2$ iterations
of block renormalization.}
\end{figure}


\begin{thebibliography}{10}
\bibitem[16]{MRT comment} The parameter $W$ in the MRT studies \cite{MRT noise measurements}
is noise in terms of energy and $WL/\phi_{0}$ should give the rough
magnitude of magnetic flux noise.
\end{thebibliography}
\end{document}